\documentclass[aps,prd,reprint,nofootinbib,longbibliography,showkeys]{revtex4-2}

\pdfoutput=1

\usepackage{blindtext}
\usepackage{mathtools}
\usepackage{xcolor}

\usepackage{dsfont}
\usepackage{siunitx}

\usepackage{mathrsfs}
\usepackage{braket}

\usepackage[utf8]{inputenc}
\usepackage[english]{babel}
\usepackage{hyperref}

\hypersetup{
    colorlinks=true,
    linkcolor=blue,
    filecolor=blue,      
    urlcolor=blue,
    citecolor=blue,
}
\usepackage{color,colortbl}

\usepackage{tensind}
\tensordelimiter{?}

\usepackage{graphicx}
\usepackage{hyperref}
\DeclareGraphicsExtensions{.bmp,.png,.jpg,.pdf}
\usepackage{amsmath}
\usepackage{courier}
\usepackage{verbatim}
\usepackage{amssymb}
\usepackage{amsfonts}
\usepackage{natbib}
\usepackage{caption}
\usepackage{subcaption}
\usepackage{csquotes}

\usepackage{float}
\usepackage{placeins}

\usepackage{booktabs}
\usepackage{siunitx}

\newcommand{\be}{\begin{equation}}
\newcommand{\ee}{\end{equation}}

\usepackage[dvipsnames]{xcolor}

\begin{document}

\title{Cosmological forecast from the full-sky angular power spectrum and bispectrum of 21cm intensity mapping}

\author{Rodrigo F. Pinheiro$^{1}$, André A. Costa$^{2}$, Yu Sang$^{3}$}
\email{rfpinheiro@cbpf.br}
\email{andre.alencar@academico.ufpb.br}
\email{sangyu@yzu.edu.cn}

\affiliation{$^1$ Centro Brasileiro de Pesquisas F\'{\i}sicas - CBPF, Rua Dr. Xavier Sigaud 150, Urca, 22290-180, Rio de Janeiro, RJ, Brazil\\
$^2$ Departamento de F\'{i}sica, Universidade Federal da Para\'{i}ba, Caixa Postal 5008, Jo\~{a}o Pessoa 58051-900, Para\'{i}ba, Brazil\\
$^3$ Center for Gravitation and Cosmology, College of Physical Science and Technology, Yangzhou University, Yangzhou 225009, China
}

\begin{abstract}

We compute the full-sky angular power spectrum and bispectrum, along with their Fisher matrices, to forecast constraints on cosmological parameters for the BINGO and SKA1-MID Band 2 radio telescopes. This represents the first forecast analysis using the full-sky relativistic bispectrum in redshift space for these surveys. Our results show that the second-order velocity contribution, often neglected under the Limber approximation, accounts for approximately $24\%$ of the total signal at low redshifts, indicating that it must be included for accurate modeling. Using these forecasts, we find that while the bispectrum provides constraints comparable to the angular power spectrum for $\Lambda$CDM and ${\rm w}$CDM models, it becomes a powerful probe of dynamical dark energy. Restricting the analysis to linear scales, we show that the inclusion of the bispectrum yields a substantial improvement in the determination of the Chevallier-Polarski-Linder (CPL) parameters. In particular, the joint analysis of the bispectrum, power spectrum, and Planck CMB data improves constraints on ${\rm w}_0$ and ${\rm w}_a$ by over $70\%$, and the Hubble parameter $h$ by approximately $60\%$. These results underscore the importance of relativistic bispectrum for breaking parameter degeneracies and probing the nature of dark energy with upcoming large-scale structure surveys.

\end{abstract}

\maketitle

\section{Introduction}
\label{sec:intro}

One of the greatest challenges in modern cosmology is to explain the origin of the accelerated expansion of the universe, which was first directly observed through observations of type Ia supernovae \cite{SupernovaSearchTeam:1998fmf,SupernovaCosmologyProject:1998vns}. The simplest theoretical explanation of the accelerated expansion comes from the cosmological constant $\Lambda$ in the Einstein equation. However, this hypothesis suffers from serious theoretical problem \cite{Weinberg:1988cp} and is also increasingly challenged by the recent baryon acoustic oscillation (BAO) measurements from Dark Energy Spectroscopic Instrument (DESI) DR2 \cite{DESI:2024mwx}, which show a preference for dynamical dark energy models.

In general, a combination of cosmic microwave background (CMB) and large-scale structure (LSS) observations is required to effectively probe the nature of dark energy. The CMB anisotropies have been measured with great precision by the Planck satellite \cite{Planck:2018vyg}, yielding stringent constraints on the cosmological parameters. However, CMB data alone are insufficient to tightly constrain dark energy parameters. Therefore, next-generation LSS galaxy surveys like DESI \cite{DESI:2016fyo, DESI:2016igz}, Euclid \cite{Amendola:2016saw, Euclid:2019clj}, and the Vera Rubin \cite{LSST:2008ijt, LSSTDarkEnergyScience:2018jkl} telescopes will play a crucial role in unraveling the properties of the accelerated expansion of the universe. 

Just as galaxy clusters and galaxies serve as tracers of the matter field, the spatial distribution of neutral hydrogen (HI) in the universe can also be used to trace the underlying matter field \cite{Pritchard:2011xb,Battye:2004re,Chang:2007xk,Loeb:2008hg,Sethi:2005gv,Visbal:2008rg}. Radio telescopes such as the BAO from Integrated Neutral Gas Observations (BINGO) \cite{Abdalla:2021nyj, Wuensche:2021dcx} and the Square Kilometer Array (SKA) \cite{Weltman:2018zrl, SKA:2018ckk} are designed to map the 21cm brightness temperature fluctuations of neutral hydrogen during the post-reionization era, a period in which dark energy significantly influences the evolution of the universe.

The majority of the cosmological information accessible to date is extracted from the two-point correlation function in spherical harmonic or Fourier space, i.e., the angular power spectrum $C_\ell(z)$ or the Fourier power spectrum $P(k)$. Many works have used the angular power spectrum to investigate cosmology \cite{Costa:2021jsk,Hall:2012wd,Chen:2019jms,Xiao:2021nmk,Ostergaard:2024brd,Song:2026mqf}. The angular power spectrum from BINGO and SKA can improve the Planck CMB constraints on dark energy cosmological parameters by up to 93$\%$ and 98$\%$ for ${\rm w}_{0}$ and ${\rm w}_{a}$, respectively, within the Chevallier-Polarski-Linder (CPL) parametrization model \cite{Costa:2021jsk}. It has been shown that the three-point correlation function can further improve the cosmological constraints derived from power spectrum and breaks parameter degeneracies \cite{Ivanov:2021kcd,Philcox:2021kcw}. For SKA, the bispectrum in Fourier space can yield parameter constraints comparable in precision to those derived from the power spectrum \cite{Karagiannis:2022ylq}, and their combination improves the 21cm power spectrum parameter constraints on ${\rm w}_{0}$ and ${\rm w}_{a}$ by approximately 24$\%$ and 20$\%$, respectively. The bispectrum contains substantial information regarding the cosmological evolution of the universe and its constituent components, making it a powerful probe for constraining deviations from standard cosmological models.

In this paper, we use the angular power spectrum and bispectrum from 21cm intensity mapping to study the capability of the future single-dish telescope to constrain cosmological parameters. Although similar issues have been explored in a previous study for SKA \cite{Karagiannis:2022ylq}, this study was conducted in the Fourier space. In contrast, our work does not adopt the flat-sky approximation but instead performs a full-sky analysis through spherical harmonics. Compared with the flat-sky approximation, the full-sky analysis avoids geometric simplifications, incorporates information across all angular scales, particularly preserving large-scale modes, and is free from edge effects which arise in the Fourier transform of a finite spatial domain. For small sky area, flat-sky approximation is reasonable and is widely used in current galaxy surveys. However, for next-generation surveys including wide-angle correlations and higher redshift slices, flat-sky approximation is insufficient and the full-sky analysis is required \cite{DiDio:2015bua,DiDio:2018unb}.

This paper is organized as follows: Section~\ref{sec:model} describes the theoretical estimation for the angular power spectrum and bispectrum signal; In Section~\ref{sec:survey}, we present the survey characteristics and instrumental noise; The Fisher matrix formalism, used in our analysis, is explained in Section~\ref{sec:fisher}; Finally, Section~\ref{sec:results} show our results and Section~\ref{sec:conclusion} has our conclusions. 

\section{21cm Signal}
\label{sec:model}

\subsection{Angular power spectrum}
As in the case of CMB, the 21cm temperature fluctuations encode primordial universe physics and its cosmological evolution. These fluctuations are described by the HI temperature contrast, which is defined as
\begin{equation}
    \Delta(\mathbf{n},z) = \frac{T_{\rm HI}(\mathbf{n},z) - \langle T_{\rm HI} \rangle(z)}{\langle T_{\rm HI} \rangle(z)},
\end{equation}
where $T_{\rm HI}$ is the brightness temperature, $\mathbf{n}$ is a unit vector that points from the source to the observer, and $z$ is the source redshift. Here $\langle \cdots \rangle$ denotes the ensemble average over all possible realizations. In actual observations, this is replaced by the angular average at fixed observed redshift $z$. Theoretically it can be calculated as
\begin{align}\label{eq:Tb_mean}
    \langle T_{\rm HI}\rangle(z) & = \frac{3(h_pc)^3\bar{n}_{\rm HI} A_{10}}{32 \pi k_B E_{21}^2 (1+z) H(z)} \,, \nonumber \\
    & = 188 \,h\,\Omega_{\rm HI}(z)\frac{(1 + z)^2}{E(z)} \, \text{mK} \,,
\end{align}
where $A_{10}$ is the coefficient of spontaneous emission, $\bar{n}_{\rm HI}$ is the average hydrogen number density at redshift $z$ in a rest frame, and $E_{21} = 5.88 \times 10^{-6} {\rm eV}$ is the energy difference between the two energy levels associated with the $\rm{HI}$ hyperfine splitting. Moreover, $E(z) = H(z)/H_0$ is the normalized Hubble parameter with $H_0 = 100h \, \rm{km \, s^{-1} \, Mpc^{-1}}$, $\Omega_{\rm{HI}}(z)$ parametrizes the $\rm{HI}$ density in units of the current critical density, and $c$, $h_p$, and $k_B$ represent the light speed, the Planck constant, and the Boltzmann constant, respectively.

To write the equations in terms of observable variables like angles and redshifts, we will now expand the temperature perturbation in the spherical harmonics
\begin{align}
    \Delta(\mathbf{n}, z) =& \sum_{\ell = 0}^{\infty} \sum_{m = -\ell}^{\ell}a_{\ell m}(z)Y_{\ell m}(\mathbf{n}), \label{DeltaSH}
\end{align}
where
\begin{align}
    \quad a_{\ell m}(z) =& \int d\Omega_{\mathbf{n}}\Delta(\mathbf{n},z)Y^{\ast}_{\ell m}(\mathbf{n}) \,. \label{alm}    
\end{align}
The angular power spectrum corresponds to the two point correlation function $\langle\Delta (\mathbf{n}, z_i)\Delta (\mathbf{n'}, z_j)\rangle$, and can be written as
\begin{equation}
\left\langle a_{\ell m}(z_i)a_{\ell 'm'}^*(z_j) \right\rangle = \delta_{\ell \ell'}\delta_{mm'} C_\ell (z_i,z_j) \,.
\end{equation}
It can be estimated theoretically from the power spectrum of primordial curvature perturbations $\mathcal{P_R}(k)$ convolved with 21cm perturbations
\begin{equation}\label{eq:cl}
C_\ell (z_i, z_j) = 4\pi \int d\ln{k} \mathcal{P_R}(k)\Delta_{T_{b,\ell}}^W(\mathbf{k}, z_i)\Delta_{T_{b,\ell}}^{W'}(\mathbf{k}, z_j)\, ,
\end{equation}
where
\begin{equation}\label{eq:DW}
\Delta_{T_{b,\ell}}^W(\mathbf{k},z) = \int_{0}^{\infty}dz \langle T_{\rm HI}\rangle(z) W(z)\Delta_{T_{b,\ell}}(\mathbf{k},z) \,.
\end{equation}

In the Poisson gauge, the scalar modes of the perturbed FLRW metric are \cite{Mukhanov:2005sc, Ma:1995ey}:
\begin{equation}
    ds^{2} = a^{2}[-(1 + 2\Psi)d\eta^{2} + (1 - 2\Phi)d\mathbf{x}^{2}]\, ,
\end{equation}
where $a$ is the scale factor, $\Psi$ is the Newton potential, and $\Phi$ is the curvature perturbation. Combining with these metric perturbations, the brightness temperature perturbations of neutral hydrogen at linear order can be calculated in Fourier space as \cite{Hall:2012wd}
\begin{eqnarray}\label{Eq:Delta_T_bl}
\Delta_{T_{b,\ell}}(\mathbf{k},z) &=&  \delta_{\rm H_{\rm I}}j_\ell(k\chi) + \frac{kv}{\mathcal{H}}j_\ell''(k\chi) + \left(\frac{1}{\mathcal{H}}\dot{\Phi}+\Psi\right)j_\ell(k\chi) \nonumber \\ 
&& - \left[\frac{1}{\mathcal{H}}\frac{d\ln(a^3\bar{n}_{\rm H_{\rm I}})}{d\eta} - \frac{\dot{\mathcal{H}}}{\mathcal{H}^2}-2\right]\Bigg[\Psi j_\ell(k\chi) \nonumber \\
&& + vj_\ell'(k\chi) + \int_{0}^{\chi}(\dot{\Psi} + \dot{\Phi})j_\ell(k\chi ')d\chi '\Bigg].
\end{eqnarray}
It includes contributions from HI density, redshift-space distortion (RSD), integrated Sachs-Wolfe (ISW) eﬀect, etc (see Ref \cite{Hall:2012wd} for details).

\subsection{Bispectrum in harmonic space}

In this work, we assume that the primordial perturbations are Gaussian. So in first order, all the information is encoded by the two point correlation $\langle\Delta^{(1)}\Delta^{(1)}\rangle$, which corresponds to the power spectrum. The three point correlation of the first order perturbation,  $\langle\Delta^{(1)}\Delta^{(1)}\Delta^{(1)}\rangle$, vanishes. We will address the discussion on bispectrum from primordial non-Gaussianity (PNG) in a future work. In this paper, instead, we focus on the bispectrum from the non-linear gravitational clustering.
To calculate the relativistic bispectrum we need to go to second order perturbation equations of the HI temperature contrast. So, up to second order we can write  
\begin{equation}
    \Delta = \Delta^{(1)} + \Delta^{(2)},
\end{equation}
where $\Delta^{(1)}$ is the linear order perturbation and $\Delta^{(2)}$ is the second order perturbation of HI temperature contrast. The linear order perturbation $\Delta^{(1)}$ is exact the one we use to calculate angular power spectrum, written as Eq.~\eqref{Eq:Delta_T_bl} in Fourier space. However, for the calculation of the power spectrum and the bispectrum, we only consider the dominated terms, i.e., the density and RSD terms, written in real space as \cite{Hall:2012wd} 
\begin{equation}
    \Delta^{(1)} = b_{1}\delta^{(1)} + \mathcal{H}^{-1}\partial^{2}_{r}v^{(1)},
\end{equation}
where $\delta^{(1)}$ is the matter field first order perturbation, $v^{(1)}$ is the velocity field first order perturbation, $r$ is the comoving line-of-sight distance, and $b_{1}$ is the linear HI bias. We neglect terms that are proportional to $\mathcal{H}/k$ because they are negligible compared to those in the equation.

Second order perturbation of the brightness temperature  has many more terms. The important ones are the following \cite{Umeh:2015gza}
\begin{align}
    \Delta^{(2)}(\mathbf{n},z) =& b_{1}\delta^{(2)} +\frac{1}{2}b_{2}(\delta^{(1)})^{2} + b_{s}s^{2} + \mathcal{H}^{-1}\partial^{2}_{r}v^{(2)} \nonumber\\
      &+ \mathcal{H}^{-2}[(\partial^{2}_{r}v^{(1)})^{2} + \partial_{r}v^{(1)}\partial^{3}_{r}v^{(1)}] \nonumber \\
      &+ \mathcal{H}^{-1}[\partial_{r}v^{(1)}\partial_{r}\delta^{(1)} + \partial^{2}_{r}v^{(1)}\delta^{(1)}],
\end{align}
where $\delta^{(2)}$ and $v^{(2)}$ are the matter density and the second-order velocity field perturbations, respectively. In this equation, we neglect lensing terms. It was shown that the brightness temperature is not affected by lensing in first order perturbations \cite{Jalilvand:2018ikk}. In addition, when we deal with bin correlations of the temperature map at the same redshift, lensing effects are negligible. For this reason, we will only consider redshift bin auto-correlations in the Fisher matrix analysis.    


The parameters $b_{1}$, $b_{2}$ and $b_{s}$ correspond to the local model clustering bias, following \cite{Desjacques:2016bnm}:
\begin{equation}
    \delta_{\rm HI} = b_{1}\delta + \frac{1}{2}b_{2}\delta^{2} + b_{s}s^{2},  
\end{equation}
where $\delta = \delta^{(1)} + \delta^{(2)}$ is the matter overdensity in the comoving gauge, and the bias coefficients are assumed to be scale-independent. The tidal field is defined as:
\begin{equation}
    s^{2} = s_{ij}s^{ij}, \quad s_{ij} = \frac{2}{3\Omega_{M}\mathcal{H}^{2}}\partial_{i}\partial_{j}\Phi - \frac{1}{3}\delta_{ij}\delta,
\end{equation}
and $b_{s}$ represents the tidal bias. Under the assumption that there is no tidal bias at the time of galaxy formation, $b_{s}$ takes the simple form:
\begin{equation}
    b_{s}(z) = -\frac{2}{7}\bar{b}_{s}(b_{1}(z) - 1),
    \label{bs}
\end{equation}
where $\bar{b}_{s}$ is an amplitude parameter with a fiducial value of one. From numerical simulations, the bias coefficients can be modeled as redshift-dependent polynomial fits \cite{Umeh:2015gza}:
\begin{equation}
    b_{1}(z) = \bar{b}_1(0.9 + 0.4z),
    \label{b1}
\end{equation}
\begin{align}
    \nonumber b_{2}(z) =& \bar{b}_{2}(-0.704172 - 0.207993z + \\
     &+ 0.183023z^2 - 0.00771288z^3),
     \label{b2}
\end{align}
where  $\bar{b}_1$ and $\bar{b}_2$ are also amplitude parameters with fiducial values set to one. 

If the brightness temperature map is Gaussian, its three-point correlation function vanishes. Consequently, the bispectrum — defined as the Fourier transform of the three-point correlation function — becomes a powerful tool for probing non-Gaussian features in 21 cm maps. These non-Gaussianities can arise from two main sources: primordial and relativistic effects. The primordial contribution originates from the physics of the early Universe, for instance from interactions of the inflaton field with other fields during inflation \cite{Maldacena:2002vr, Bartolo:2004if}. The relativistic contribution, on the other hand, is generated by the non-linear gravitational clustering responsible for the formation of cosmic structures such as galaxies and galaxy clusters \cite{Kehagias:2015tda, DiDio:2016gpd}.

The bispectrum is given by the ensemble average of the product of the temperature perturbation at three points: 
\begin{multline}
    B(\mathbf{n_{1}},\mathbf{n_{2}},\mathbf{n_{3}}, z_{1}, z_{2}, z_{3}) \equiv  \\
    \langle\Delta(\mathbf{n_{1}},z_{1}) \Delta(\mathbf{n_{2}},z_{2})\Delta(\mathbf{n_{3}},z_{3})\rangle.
     \label{bispectrum1}
\end{multline}
Assuming Gaussian initial conditions generated during inflation, the leading non-vanishing contribution to the bispectrum arises at second order and is given by
\begin{multline}
    B_{\rm tree}(\mathbf{n_{1}},\mathbf{n_{2}},\mathbf{n_{3}}, z_{1}, z_{2}, z_{3}) \equiv \\ 
     \langle\Delta^{(2)}(\mathbf{n_{1}},z_{1})\Delta^{(1)}
    (\mathbf{n_{2}},z_{2})\Delta^{(1)}(\mathbf{n_{3}},z_{3})\rangle + \circlearrowleft,    
\end{multline}
where $\circlearrowleft$ denotes the two additional terms required to complete the cyclic set. In each term, the second-order perturbation is shifted to a different coordinate pair - $({\bf n_{2}}, z_{2})$ then $({\bf n_{3}}, z_{3})$ - ensuring that the final bispectrum is symmetric and accounts for the non-linear contribution at every position in the configuration.

Expanding the bispectrum in terms of spherical harmonics as we did for the power spectrum in Eq.~\eqref{DeltaSH}, we obtain
\begin{multline}
    B(\mathbf{n_{1}},\mathbf{n_{2}},\mathbf{n_{3}}, z_{1}, z_{2}, z_{3}) = \sum_{\substack{\ell_{1},\ell_{2},\ell_{3} \\ m_{1},m_{2},m_{3}}}B_{\ell_{1}\ell_{2}\ell_{3}}^{m_{1} m_{2} m_{3}}(z_{1},z_{2},z_{3})  \\
    \times Y_{\ell_{1} m_{1}}(\mathbf{n_{1}})Y_{\ell_{2} m_{2}}(\mathbf{n_{2}})Y_{\ell_{3} m_{3}}(\mathbf{n_{3}}),
\end{multline}   
where $B_{\ell_{1}\ell_{2}\ell_{3}}^{m_{1} m_{2} m_{3}}(z_{1},z_{2},z_{3})$ are the bispectrum expansion coefficients. By substituting Eq.~\eqref{DeltaSH} into Eq.~\eqref{bispectrum1}, they can be written in terms of the $a_{\ell m}(z)$ coefficients:
\begin{align}
   B_{\ell_{1}\ell_{2}\ell_{3}}^{m_{1} m_{2} m_{3}}(z_{1},z_{2},z_{3}) =  \langle a_{\ell_{1}m_{1}}(z_{1})a_{\ell_{2}m_{2}}(z_{2})a_{\ell_{3}m_{3}}(z_{3})\rangle \,. 
\end{align} 
Now, substituting Eq.~\eqref{alm} into the above equation, we can write the bispectrum coefficients as
\begin{equation}
    B_{\ell_{1}\ell_{2}\ell_{3}}^{m_{1} m_{2} m_{3}}(z_{1},z_{2},z_{3}) = \mathcal{G}_{\ell_{1}\ell_{2}\ell_{3}}^{m_{1} m_{2} m_{3}} b_{\ell_{1}\ell_{2}\ell_{3}}(z_{1},z_{2},z_{3}) \,,
    \label{eq:bispectrum_coefficients}
\end{equation}
where $\mathcal{G}_{\ell_{1}\ell_{2}\ell_{3}}^{m_{1} m_{2} m_{3}}$ is the Gaunt integral, defined as the integral of the spherical harmonic product in three points, which can be expressed in terms of the Wigner 3j symbols
\begin{multline}
    \mathcal{G}_{\ell_{1}\ell_{2}\ell_{3}}^{m_{1} m_{2} m_{3}} =  \int d\Omega Y_{\ell_{1} m_{1}}(\mathbf{n})Y_{\ell_{2} m_{2}}(\mathbf{n})Y_{\ell_{3} m_{3}}(\mathbf{n}) \\
    = \begin{pmatrix} \ell_{1} & \ell_{2} & \ell_{3} \\ 0 & 0 & 0 \end{pmatrix} \begin{pmatrix} \ell_{1} & \ell_{2} & \ell_{3} \\ m_{1} & m_{2} & m_{3} \end{pmatrix}  \\
     \times \sqrt{\frac{(2\ell_{1} + 1)(2\ell_{2} + 1)(2\ell_{3} + 1)}{4\pi}} \,.
\end{multline}
The Gaunt integral accounts for all angular dependence of the bispectrum coefficients, Eq.~\eqref{eq:bispectrum_coefficients}, whereas the cosmological dependence is fully described by the reduced bispectrum $b_{\ell_{1}\ell_{2}\ell_{3}}(z_{1},z_{2},z_{3})$.

The 21cm reduced bispectrum has the following six terms \cite{Durrer:2020orn}:
\begin{align}
    \label{eq:reduced_bispectrum} 
    b_{\ell_{1}\ell_{2}\ell_{3}}(z_{1},z_{2},z_{3}) & = b^{\delta^{(2)}}_{\ell_{1}\ell_{2}\ell_{3}}(z_{1},z_{2},z_{3}) + b^{v^{(2)^{\prime}}}_{\ell_{1}\ell_{2}\ell_{3}}(z_{1},z_{2},z_{3}) \nonumber \\
    & + b^{\delta v^{\prime}}_{\ell_{1}\ell_{2}\ell_{3}}(z_{1},z_{2},z_{3}) + b^{v^{\prime 2}}_{\ell_{1}\ell_{2}\ell_{3}}(z_{1},z_{2},z_{3}) \nonumber \\
    & + b^{\delta^{\prime} v}_{\ell_{1}\ell_{2}\ell_{3}}(z_{1},z_{2},z_{3}) + b^{v^{\prime\prime}v}_{\ell_{1}\ell_{2}\ell_{3}}(z_{1},z_{2},z_{3})\,,
\end{align}
where the first and the second terms are related to the density and velocity second-order perturbations, respectively. The other terms come from quadratic combinations of density and velocity first order perturbations.  

The reduced bispectrum can be written in terms of the generalized power spectrum \cite{DiDio:2015bua}:
\begin{equation}
    {}^{n}c^{AB}_{\ell \ell^{\prime}}(z_{1}, z_{2}) = i^{\ell - \ell^{\prime}}4\pi\int\frac{dk}{k}k^{n}\mathcal{P}_{R}(k)\Delta^{A}_{\ell}(k,r_{1})\Delta^{B}_{\ell^{\prime}}(k,r_{2})\,,
    \label{eq:g_ps}
\end{equation}
where $\mathcal{P}_{R}(k)$ is the dimensionless curvature primordial power spectrum, $\Delta^{A}_{\ell}(k,r)$ is the angular transfer function, $A$ and $B$ are labels for the following perturbation it makes reference: 
\begin{equation}
\begin{split}
    \Delta_{\ell}^{\Delta}(k,r) &= b_{1}(r)T_{\delta}(k,r)j_{\ell}(kr) + \frac{k}{\mathcal{H}}T_{V}(k,r)j^{\prime \prime}_{\ell}(kr), \\
    \Delta_{\ell}^{\delta}(k,r) &= T_{\delta}(k,r)j_{\ell}(kr), \\
    \Delta_{\ell}^{\delta^{\prime}}(k,r) &= \frac{k}{\mathcal{H}}T_{\delta}(k,r)j^{\prime}_{\ell}(kr), \\
    \Delta_{\ell}^{v}(k,r) &= T_{V}(k,r)j^{\prime}_{\ell}(kr), \\
    \Delta_{\ell}^{v^{\prime}}(k,r) &= \frac{k}{\mathcal{H}}T_{V}(k,r)j^{\prime \prime}_{\ell}(kr), \\
    \Delta_{\ell}^{v^{\prime \prime}}(k,r) &= \left(\frac{k}{\mathcal{H}}\right)^{2}T_{V}(k,r)j^{\prime \prime \prime}_{\ell}(kr) \,.
\end{split}
\end{equation}
$T_{\delta}(k,r)$ is the linear transfer function for density, and $T_{V}(k,r) = -(\mathcal{H}/k)f(z)T_{\delta}(k,r)$ is the corresponding linear transfer function for velocity. $f(z)$ is the growth rate, and $\mathcal{H}$ is the comoving Hubble factor.

The second-order density term can be split into three terms: monopole, dipole, and quadrupole, as written below
\begin{align}
b^{\delta^{(2)}}_{\ell_{1}\ell_{2}\ell_{3}}(z_{1},z_{2},z_{3}) & = \left(b_{1}(z_{1}) + \frac{21}{34}b_{2}(z_{1})\right) b_{\ell_{1} \ell_{2} \ell_{3}}^{\delta 0}(z_{1},z_{2},z_{3}) \nonumber \\
    & + b_{1}(z_{1})b_{\ell_{1} \ell_{2} \ell_{3}}^{\delta 1}(z_{1},z_{2},z_{3}) \nonumber \\ 
    & + \left(b_{1}(z_{1}) + \frac{7}{2}b_{s}(z_{1})\right)b_{\ell_{1} \ell_{2} \ell_{3}}^{\delta 2}(z_{1},z_{2},z_{3}) \nonumber \\
    & + \circlearrowleft \,,
\end{align}
where the monopole is
\begin{equation}
    b^{\delta 0}_{\ell_{1} \ell_{2} \ell_{3}}(z_{1},z_{2},z_{3}) = \frac{34}{21}c_{\ell_{1}}^{\delta \Delta}(z_{1},z_{3})c_{\ell_{2}}^{\delta \Delta}(z_{2},z_{3}) \,, 
\end{equation}  
the dipole is
\begin{align}
     b^{\delta 1}_{\ell_{1} \ell_{2} \ell_{3}}(z_{1},z_{2},z_{3}) & = \frac{(g_{\ell_{1} \ell_{2} \ell_{3}})^{-1}}{16 \pi^{2}}\sum_{\ell^{\prime} \ell^{\prime \prime}}(2\ell^{\prime} + 1)(2\ell^{\prime \prime} + 1) \nonumber \\
     & \times Q^{\ell_{1} \ell_{2} \ell_{3}}_{1 \ell^{\prime} \ell^{\prime \prime}}[{}^{1}c^{\delta \Delta}_{\ell^{\prime \prime} \ell_{2}}(z_{3}, z_{1})^{-1}c^{\delta \Delta}_{\ell^{\prime} \ell_{3}}(z_{3}, z_{2}) + \nonumber \\
     & + {}^{-1}c^{\delta \Delta}_{\ell^{\prime \prime} \ell_{2}}(z_{3}, z_{1})^{1}c^{\delta \Delta}_{\ell^{\prime} \ell_{3}}(z_{3}, z_{2})] \,,
\end{align}
and the quadrupole is
\begin{align}
    b^{\delta 2}_{\ell_{1} \ell_{2} \ell_{3}}(z_{1},z_{2},z_{3}) & = \frac{(g_{\ell_{1} \ell_{2} \ell_{3}})^{-1}}{42 \pi^{2}}\sum_{\ell^{\prime} \ell^{\prime \prime}}(2\ell^{\prime} + 1)(2\ell^{\prime \prime} + 1) \nonumber \\
    & \times Q^{\ell_{1} \ell_{2} \ell_{3}}_{2 \ell^{\prime} \ell^{\prime \prime}}c^{\delta \Delta}_{\ell^{\prime \prime} \ell_{2}}(z_{3}, z_{1})c^{\delta \Delta}_{\ell^{\prime} \ell_{3}}(z_{3}, z_{2}) \,.
\end{align}
The geometrical factors $g_{\ell_{1} \ell_{2} \ell_{3}}$ and $Q^{\ell_{1} \ell_{2} \ell_{3}}_{\ell \ell^{\prime} \ell^{\prime \prime}}$ are defined in Appendix C of \cite{DiDio:2018unb}, the main reference we are following for the bispectrum calculation.

The second-order velocity contribution cannot be written entirely in terms of the generalized power spectrum, although part of it can. The separable and non-separable components of $v^{(2)^{\prime}}$ are given in  Appendix A of \cite{DiDio:2018unb}. 

The four remaining reduced bispectrum contributions in Eq.~\eqref{eq:reduced_bispectrum} can be written as:
\begin{align}
    b^{\delta v^{\prime}}_{\ell_{1}\ell_{2}\ell_{3}}(z_{1},z_{2},z_{3}) & = c^{\delta_{g} \Delta}_{\ell_{2}}(z_{1},z_{2})c^{v^{\prime} \Delta}_{\ell_{3}}(z_{1},z_{3}) +\\ 
    & + c^{v^{\prime} \Delta}_{\ell_{2}}(z_{1},z_{2})c^{\delta_{g} \Delta}_{\ell_{3}}(z_{1},z_{3}) + \circlearrowleft \,, \nonumber \\
    b^{v^{\prime 2}}_{\ell_{1}\ell_{2}\ell_{3}}(z_{1},z_{2},z_{3}) & = 2c^{v^{\prime} \Delta}_{\ell_{2}}(z_{1},z_{2})c^{v^{\prime} \Delta}_{\ell_{3}}(z_{1},z_{3}) + \circlearrowleft \,, \\
    b^{\delta^{\prime} v}_{\ell_{1}\ell_{2}\ell_{3}}(z_{1},z_{2},z_{3}) & = b_{1}(z_{1})c^{\delta^{\prime} \Delta}_{\ell_{2}}(z_{1},z_{2})c^{v \Delta}_{\ell_{3}}(z_{1},z_{3}) +\\ 
    & + b_{1}(z_{1})c^{v \Delta}_{\ell_{2}}(z_{1},z_{2})c^{\delta^{\prime} \Delta}_{\ell_{3}}(z_{1},z_{3}) + \circlearrowleft \,, \nonumber \\
    b^{v^{\prime \prime} v}_{\ell_{1}\ell_{2}\ell_{3}}(z_{1},z_{2},z_{3}) & = c^{v^{\prime \prime} \Delta}_{\ell_{2}}(z_{1},z_{2})c^{v \Delta}_{\ell_{3}}(z_{1},z_{3}) +\\ 
    & + c^{v \Delta}_{\ell_{2}}(z_{1},z_{2})c^{v^{\prime \prime} \Delta}_{\ell_{3}}(z_{1},z_{3}) + \circlearrowleft \,. \nonumber
\end{align}

The observable bispectrum is usually considered in its angle-averaged form, which is related to the reduced bispectrum as \cite{DiDio:2015bua}
\begin{align}
    B_{\ell_{1}\ell_{2}\ell_{3}} =& \sqrt{\frac{(2\ell_{1}+1)(2\ell_{2}+1)(2\ell_{3}+1)}{4 \pi}}  \nonumber \\ 
    \times &\begin{pmatrix} \ell_{1} & \ell_{2} & \ell_{3} \\ 0 & 0 & 0 \end{pmatrix} b_{\ell_{1}\ell_{2}\ell_{3}} \,.
\end{align}

\begin{figure*}
	\centering
	\begin{subfigure}{0.32\linewidth}
		\includegraphics[width=\linewidth]{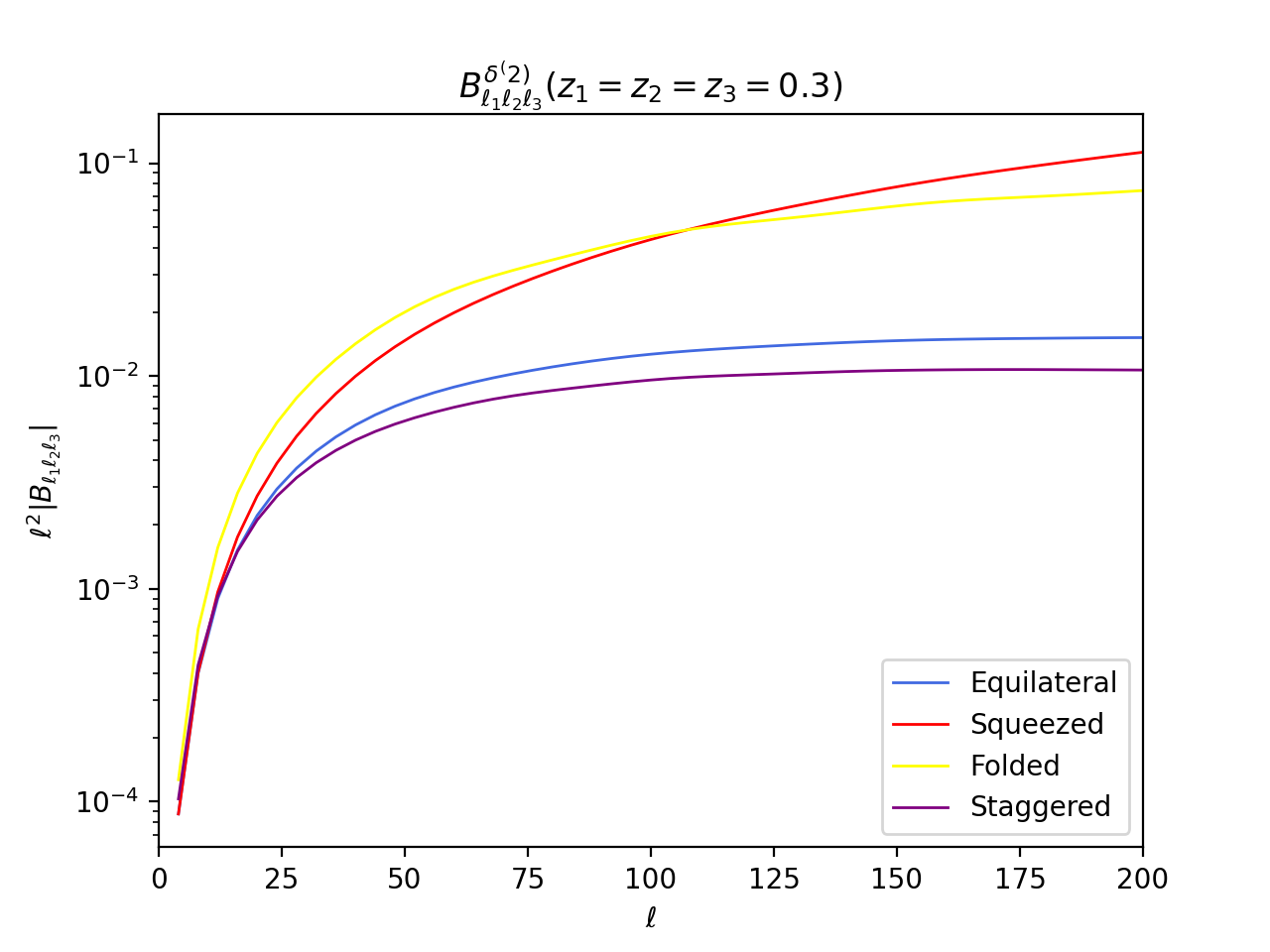}
		\label{fig:subfigA}
	\end{subfigure}
    \begin{subfigure}{0.32\linewidth}
		\includegraphics[width=\linewidth]{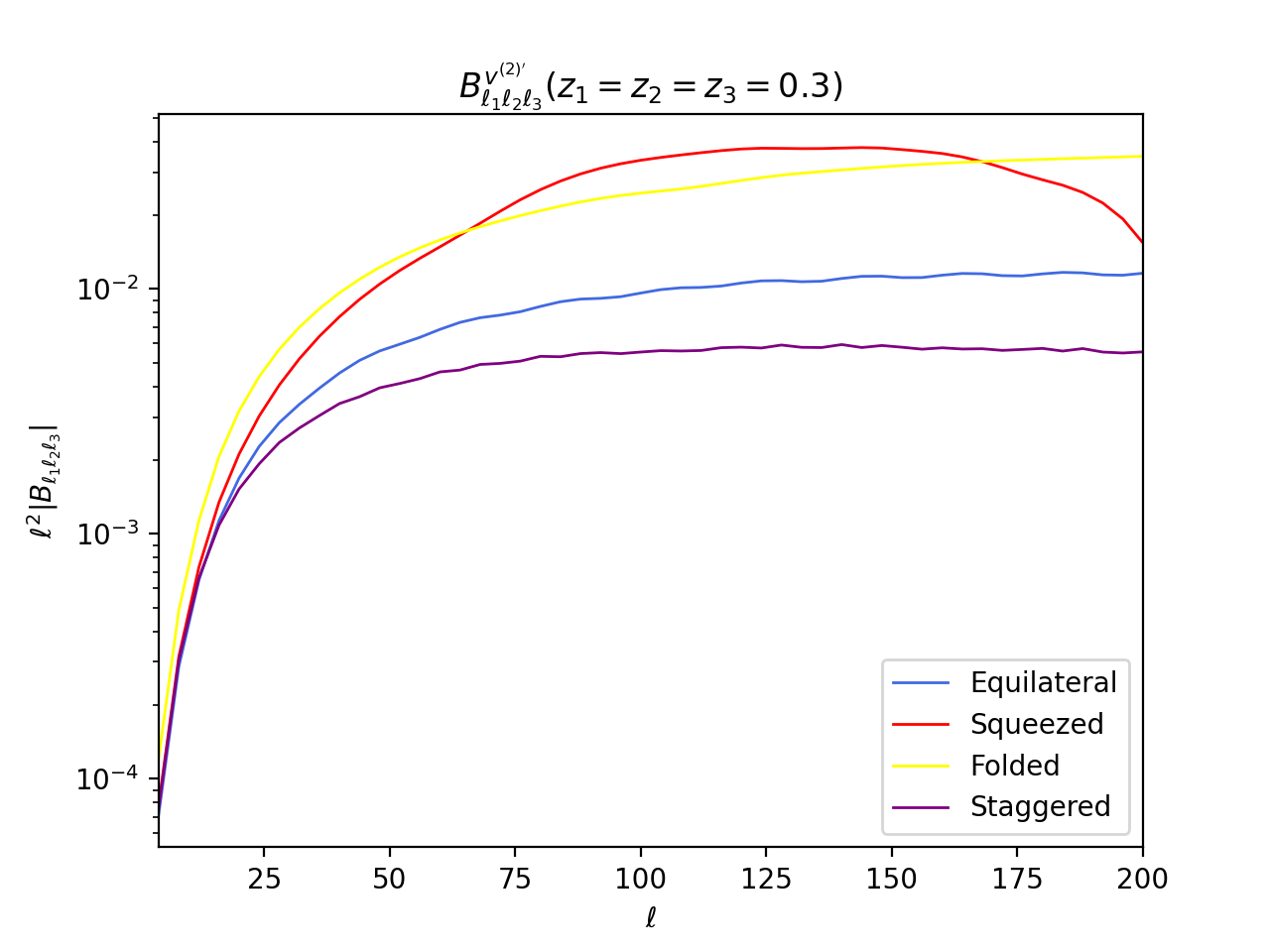}
		\label{fig:subfigB}
	\end{subfigure}
	\begin{subfigure}{0.32\linewidth}
		\includegraphics[width=\linewidth]{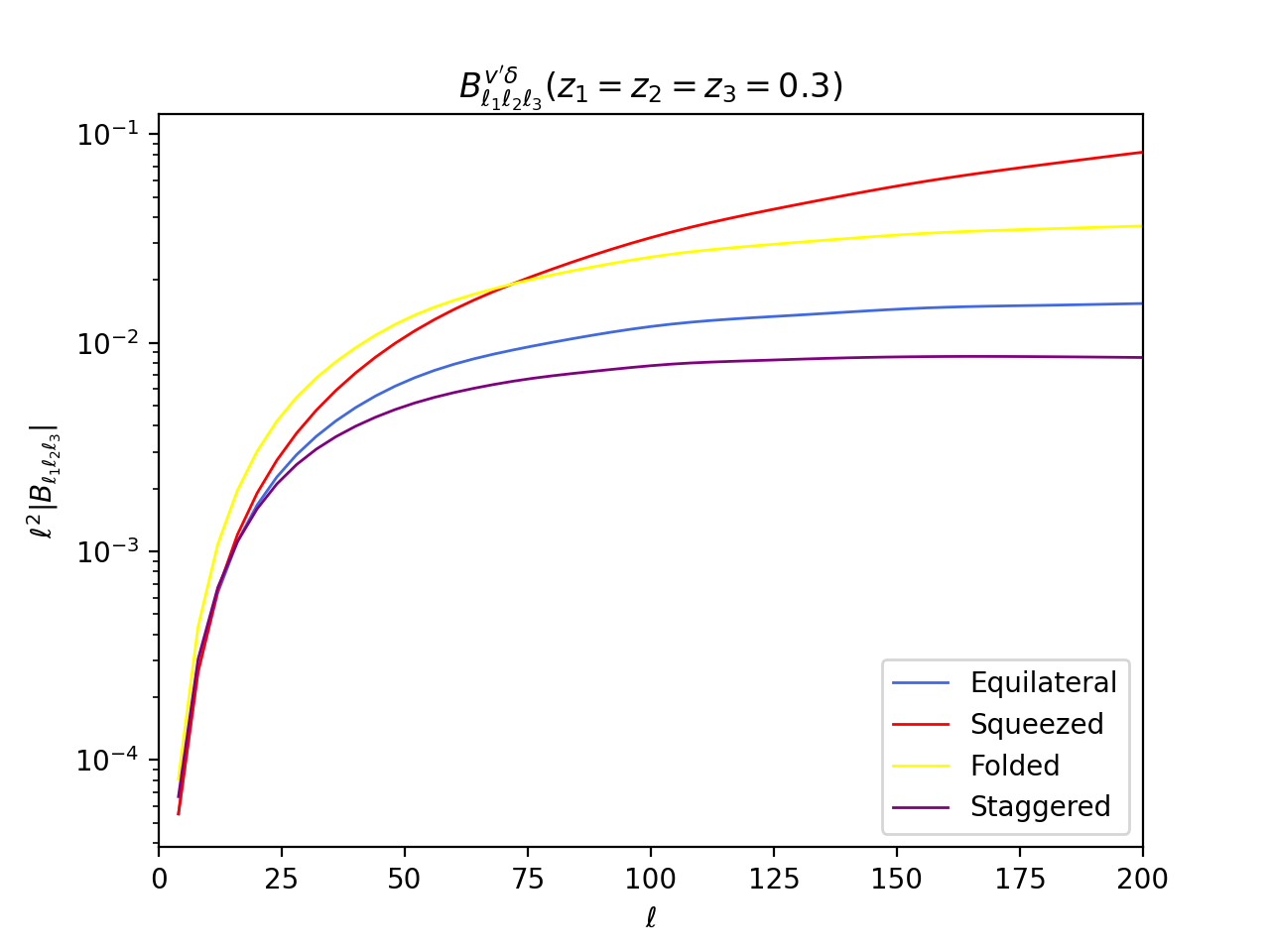}
		\label{fig:subfigC}
	\end{subfigure}
	\begin{subfigure}{0.32\linewidth}
	    \includegraphics[width=\linewidth]{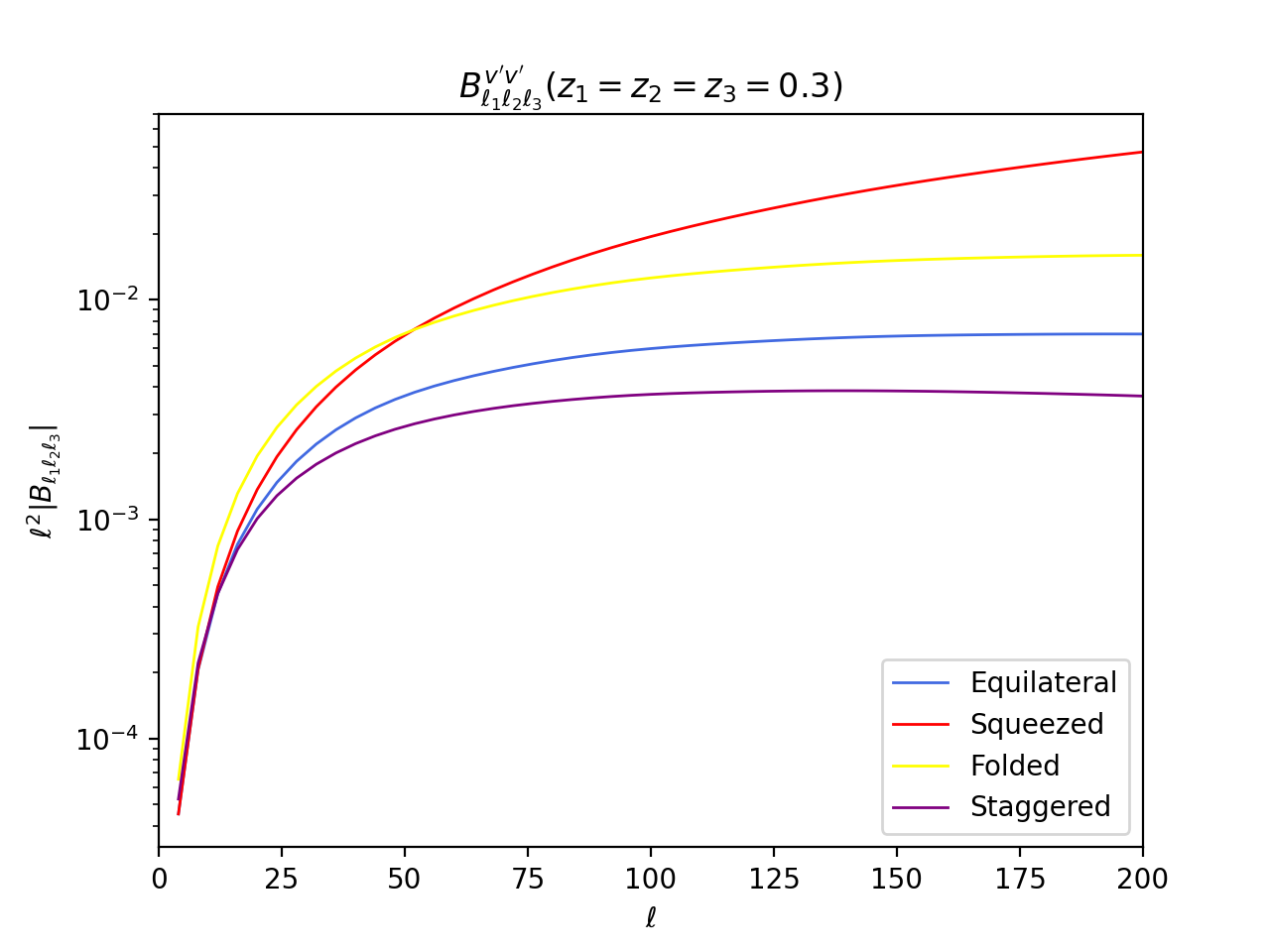}
	    \label{fig:subfigD}
    \end{subfigure}
    \begin{subfigure}{0.32\linewidth}
	    \includegraphics[width=\linewidth]{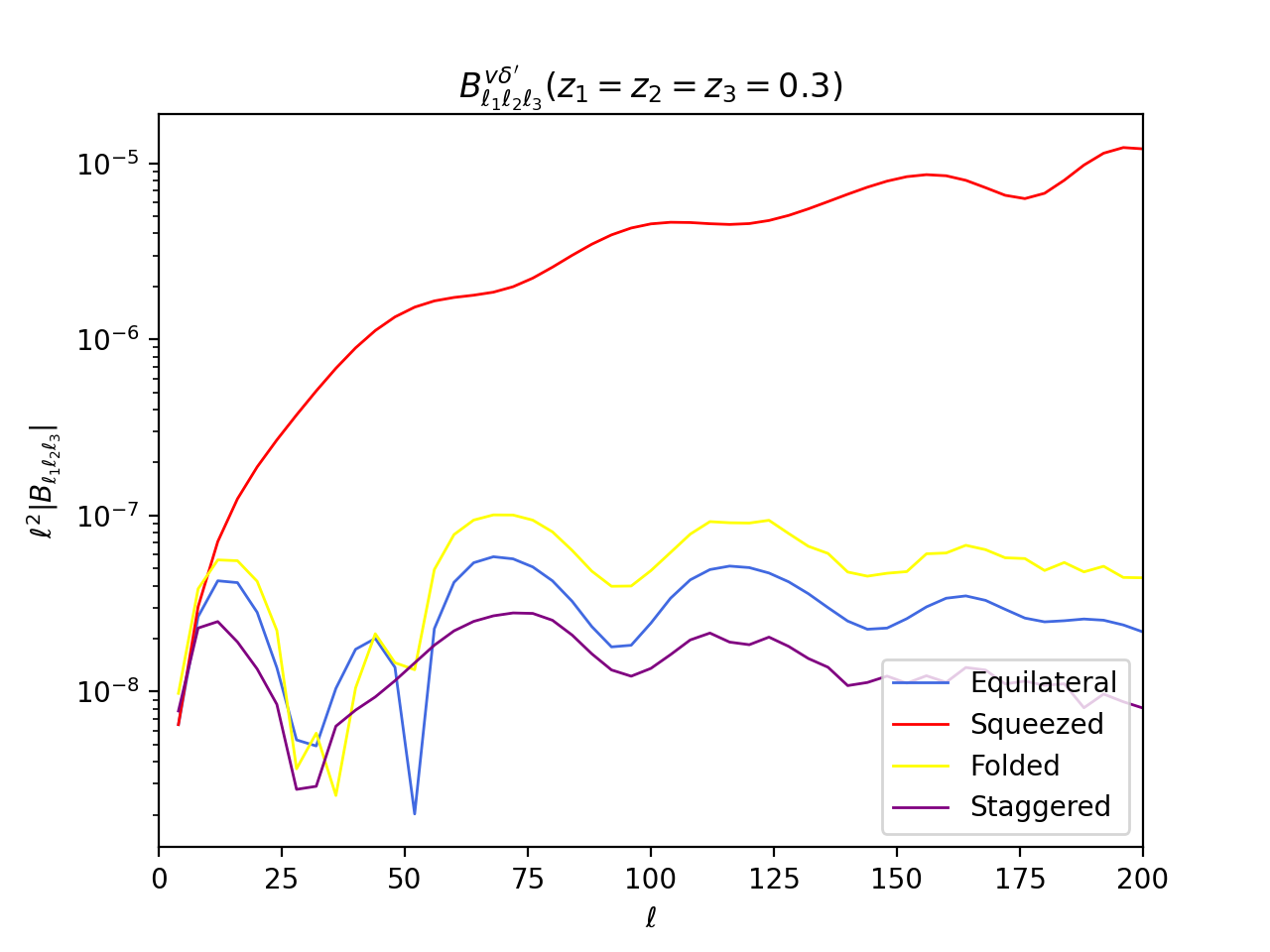}
	    \label{fig:subfigE}
    \end{subfigure}
    \begin{subfigure}{0.32\linewidth}
	    \includegraphics[width=\linewidth]{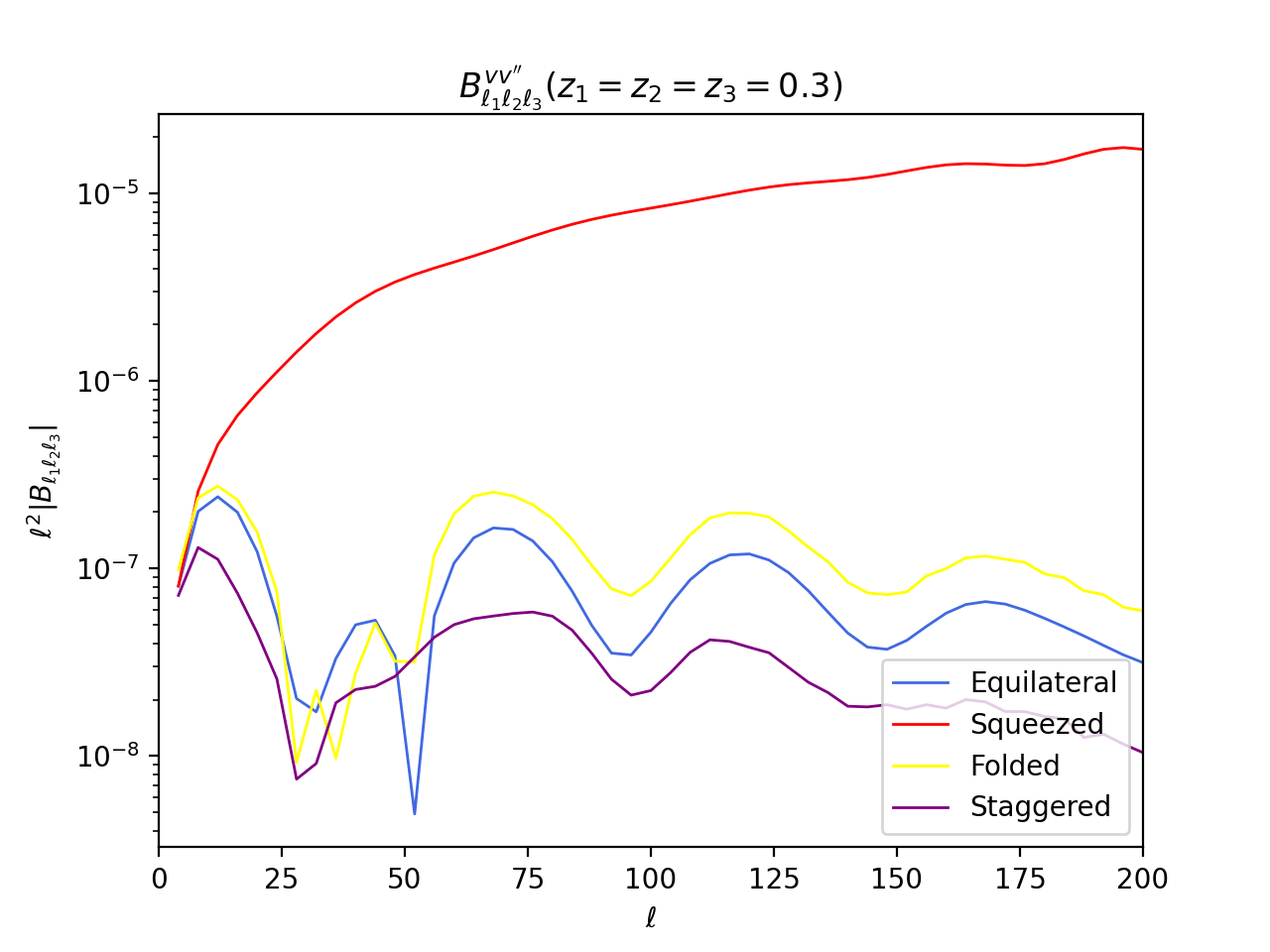}
	    \label{fig:subfigF}
    \end{subfigure}     
	\caption{The six components of the angle-averaged 21cm bispectrum at $z = 0.3$, shown for equilateral, squeezed, folded, and staggered triangle configurations. The second-order velocity term ($v^{(2)^{\prime}}$) provides a significant contribution, while the final two components remain negligible in all configurations.}
	\label{fig:avrgb}
\end{figure*}

Figure~\ref{fig:avrgb} shows the comparison of the angle-averaged bispectrum's terms for four configurations at redshift $z = 0.3$: equilateral ($\ell_{1} = \ell_{2} = \ell_{3}$), squeezed ($\ell_{1} = 4, \ell_{2} = \ell_{3}$), folded ($\ell_{1} = \ell_{2} = \ell_{3}/2$) and staggered ($\ell_{1} = \ell, \ell_{2} = 1.5\ell, \ell_{3} = 2\ell$). We can see that the last two terms ($B^{v\delta^{\prime}}_{\ell_{1}\ell_{2}\ell_{3}}$ and $B^{vv^{\prime \prime} }_{\ell_{1}\ell_{2}\ell_{3}}$ ) are negligible in relation to the others and we will ignore them in the Fisher matrix analysis. We can also realize that the second-order velocity term $B^{v^{(2)^{\prime}}}_{\ell_{1}\ell_{2}\ell_{3}}$ is critical. This component is very difficult to calculate numerically, and we will deal with it in the subsection \ref{subsec:v2}.    

Observations are typically performed over a finite redshift range rather than at a single redshift. To account for this, we define the $z$-binned angle-averaged bispectrum, $\bar{B}_{\ell_1\ell_2\ell_3}(z_1,z_2,z_3)$, by convolving the underlying bispectrum with a normalized window function $W(z,z^{\prime})$ centered at $z$:
\begin{multline}
    \bar{B}_{\ell_{1}\ell_{2}\ell_{3}}(z_{1},z_{2},z_{3}) = \sqrt{\frac{(2\ell_{1}+1)(2\ell_{2}+1)(2\ell_{3}+1)}{4 \pi}}  \\ 
    \times \begin{pmatrix} \ell_{1} & \ell_{2} & \ell_{3} \\ 0 & 0 & 0 \end{pmatrix} \int dz^{\prime}_{1} dz^{\prime}_{2} dz^{\prime}_{3} W(z_{1},z^{\prime}_{1}) W(z_{2},z^{\prime}_{2}) W(z_{3},z^{\prime}_{3})  \\
    \times b_{\ell_{1}\ell_{2}\ell_{3}}(z^{\prime}_{1},z^{\prime}_{2},z^{\prime}_{3}) \,.
\end{multline}

\subsection{Estimation of the $v^{(2)^{\prime}}$ term}
\label{subsec:v2}

It was shown that the second–order velocity contribution $v^{(2)^{\prime}}$ plays a crucial role in the calculation of the relativistic bispectrum when thin redshift bins are considered \cite{Durrer:2020orn, DiDio:2018unb}. In addition, the Limber approximation was shown to be inaccurate on large angular scales ($\ell \lesssim 100$) and for narrow redshift bins ($\Delta_z \lesssim 0.1$), precisely the regime where redshift-space distortion (RSD) effects become relevant. This makes the use of the full-sky formalism, together with the explicit evaluation of the second-order velocity term, particularly important for 21cm intensity-mapping surveys, where the redshift bin width is typically small.

As discussed in detail in Ref.~\cite{DiDio:2018unb}, the contribution $v^{(2)^{\prime}}$ to the relativistic bispectrum can be decomposed into two parts. The first one can be written in terms of the generalized power spectrum Eq.~\eqref{eq:g_ps} and is therefore numerically inexpensive to evaluate. The second part, however, involves a quadruple integral and is computationally very demanding. Since the bispectrum must be computed for a large number of triangle configurations in the Fisher matrix analysis, a direct evaluation of this term becomes prohibitive. In this section, we therefore develop an approximation for the $v^{(2)^{\prime}}$ contribution based on the remaining bispectrum components, allowing us to avoid its explicit computation in the Fisher analysis.

We define the fractional factor $r_{\ell_{1}\ell_{2}\ell_{3}}$ as the contribution of the second-order velocity term $v^{(2)^{\prime}}$ relative to the total bispectrum:
\begin{equation}
    r_{\ell_{1}\ell_{2}\ell_{3}} \equiv \frac{B^{v^{(2)^{\prime}}}_{\ell_{1}\ell_{2}\ell_{3}}}{B^{\text{total}}_{\ell_{1}\ell_{2}\ell_{3}}},  
\end{equation}
where the total bispectrum is given by $B^{\text{total}}_{\ell_{1}\ell_{2}\ell_{3}} = B^{\delta^{(2)}}_{\ell_{1}\ell_{2}\ell_{3}} + B^{v^{(2)^{\prime}}}_{\ell_{1}\ell_{2}\ell_{3}} + B^{\delta v^{\prime}}_{\ell_{1}\ell_{2}\ell_{3}} + B^{v^{\prime 2}}_{\ell_{1}\ell_{2}\ell_{3}}$. This allows us to isolate the second-order velocity contribution and rewrite it in terms of the remaining components:
\begin{equation}
    B^{v^{(2)^{\prime}}}_{\ell_{1}\ell_{2}\ell_{3}} = \frac{r_{\ell_{1}\ell_{2}\ell_{3}}}{1 - r_{\ell_{1}\ell_{2}\ell_{3}}}\left(B^{\delta^{(2)}}_{\ell_{1}\ell_{2}\ell_{3}} + B^{\delta v^{\prime}}_{\ell_{1}\ell_{2}\ell_{3}} + B^{v^{\prime 2}}_{\ell_{1}\ell_{2}\ell_{3}}\right)\,, \label{eq:v2}     
\end{equation}
Using this relation, the total bispectrum can then be expressed entirely in terms of the remaining components and the fractional factor: 
\begin{equation}
    B^{\text{total}}_{\ell_{1}\ell_{2}\ell_{3}} = \left(\frac{1}{1 - r_{\ell_{1}\ell_{2}\ell_{3}}}\right)\left(B^{\delta^{(2)}}_{\ell_{1}\ell_{2}\ell_{3}} + B^{\delta v^{\prime}}_{\ell_{1}\ell_{2}\ell_{3}} + B^{v^{\prime 2}}_{\ell_{1}\ell_{2}\ell_{3}}\right) \,.   \label{eq:B_total} 
\end{equation}
With this approach, $r_{\ell_{1}\ell_{2}\ell_{3}}$ is initially computed numerically, which still requires evaluating  $v^{(2)^{\prime}}$. However, if $r_{\ell_{1}\ell_{2}\ell_{3}}$   converges to a stable value, it can be used directly in Eq.~\eqref{eq:B_total}, eliminating the need to compute $v^{(2)^{\prime}}$ explicitly. This results in a substantial speed-up of the numerical evaluation of the total bispectrum while preserving accuracy.

To determine the ratio $r_{\ell_{1}\ell_{2}\ell_{3}}$ we evaluate a subset of triangle configurations at selected redshifts. We consider the minimum and maximum redshifts relevant for the BINGO survey, $z=0.127$ and $z = 0.45$, and for SKA1-MID Band 2, $z=0.01$ and $z = 0.49$,  together with the mean redshift of the BINGO survey, $z=0.3$. For each redshift, we compute four representative triangle configurations within the linear regime, defined by $\ell < \ell^{\text{nl}}_{\text{max}}$ (see the subsection \ref{subsec:angular-scale-limits}): equilateral, squeezed, folded, and staggered.

It is convenient to replace the configuration-dependent quantity $r_{\ell_{1}\ell_{2}\ell_{3}}$ appearing in Eq.~\eqref{eq:B_total} by a single effective parameter when estimating the total bispectrum. We therefore introduce the multipole-averaged ratio
\begin{equation}
r \equiv \langle r_{\ell_{1}\ell_{2}\ell_{3}} \rangle , 
\end{equation}
where the average is taken over the selected subset of multipole configurations.

The validity of this approximation is assessed by comparing the estimated total bispectrum with the exact result. In Table~\ref{table:approximation}, we report the averaged ratio $r$, the mean fractional error $d$ in the reconstructed bispectrum, the corresponding variance $\sigma$, and the maximum deviation obtained over the sampled configurations.

\begin{table*}
\centering
\caption{Statistic for the bispectrum approximation at $z = 0.01, 0.127, 0.3, 0.45$, and $0.49$. The second column shows the nonlinear cutoff scale for each redshift; which defines the maximum multipole used in the approximation. For $z = 0.01$, corresponding to the minimum redshift of SKA1-MID band 2, the cutoff is $\ell_{\rm max}^{\rm nl} = 3 < \ell_{\rm min}^{\rm fground}$; therefore, this redshift lies effectively outside our Fisher matrix analyses. In this case, the result is obtained by extrapolating the maximum multipole to $\ell_{\rm max} = 12$, in order to include more points in the computation. The remaining columns show, respectively, the averaged fractional parameter $r$, the error, the variance and the maximum error obtained when using the bispectrum approximation instead of the full exact bispectrum. Values in parentheses correspond to $r=0.24$.}

\vspace{0.5cm}

\setlength{\tabcolsep}{10pt}

\begin{tabular}{cccccc}\toprule

$z$ & $\ell^{\text{nl}}_{\text{max}}$ & $r$ & $d(\%)$  & $\sigma (\%)$ & maximum error $(\%)$  \\

\hline

\addlinespace[0.5ex]

0.01  & 3   & 0.2300 & 1.9$\%$ (2.1$\%$)& 1.3$\%$ (1.7$\%$)& 3.7$\%$ (4.9$\%$)\\
0.127 & 39  & 0.2392 & 1.8$\%$ (1.8$\%$)& 1.5$\%$ (1.5$\%$)& 6.8$\%$ (6.9$\%$)\\
0.3   & 99  & 0.2448 & 2.0$\%$ (2.1$\%$)& 1.6$\%$ (1.6$\%$)& 7.6$\%$ (6.9$\%$)\\
0.45  & 153 & 0.2429 & 2.0$\%$ (2.1$\%$)& 1.6$\%$ (1.6$\%$)& 7.6$\%$ (7.2$\%$)\\
0.49  & 168 & 0.2402 & 2.2$\%$ (2.2$\%$)& 1.8$\%$ (1.8$\%$)& 8.6$\%$ (8.5$\%$)\\

\addlinespace[0.5ex]

\bottomrule

\end{tabular}

\label{table:approximation}
\end{table*}

Using the estimated form of the $v^{(2)^{\prime}}$ contribution, Eq. ~\eqref{eq:v2}, the mean error in the total bispectrum signal never exceeds 2.2$\%$, with a variance of 1.8$\%$, indicating that the typical error remains below 4$\%$. In rare cases the deviation is larger, but it never exceeds $9\%$.
We find that the averaged ratio takes a value $r \approx 0.24$, implying that approximately $24\%$ of the total 21cm bispectrum signal at low redshift ($z \lesssim 0.5$) is sourced by the second-order velocity contribution, indicating that it must be included for accurate modeling.

\begin{figure*}
	\centering
	\begin{subfigure}{0.32\linewidth}
		\includegraphics[width=\linewidth]{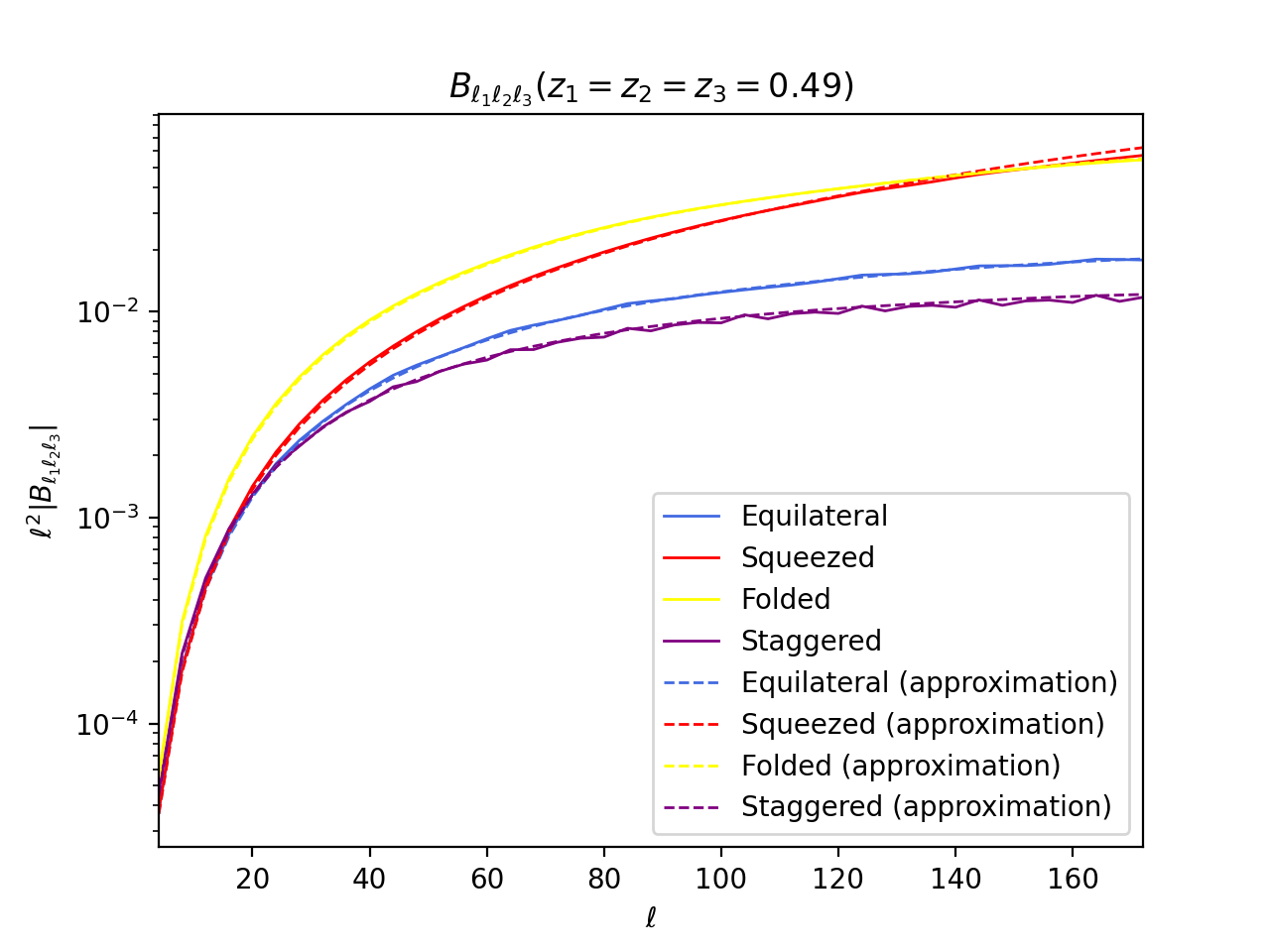}
		\label{fig:subfiga}
	\end{subfigure}
    \begin{subfigure}{0.32\linewidth}
		\includegraphics[width=\linewidth]{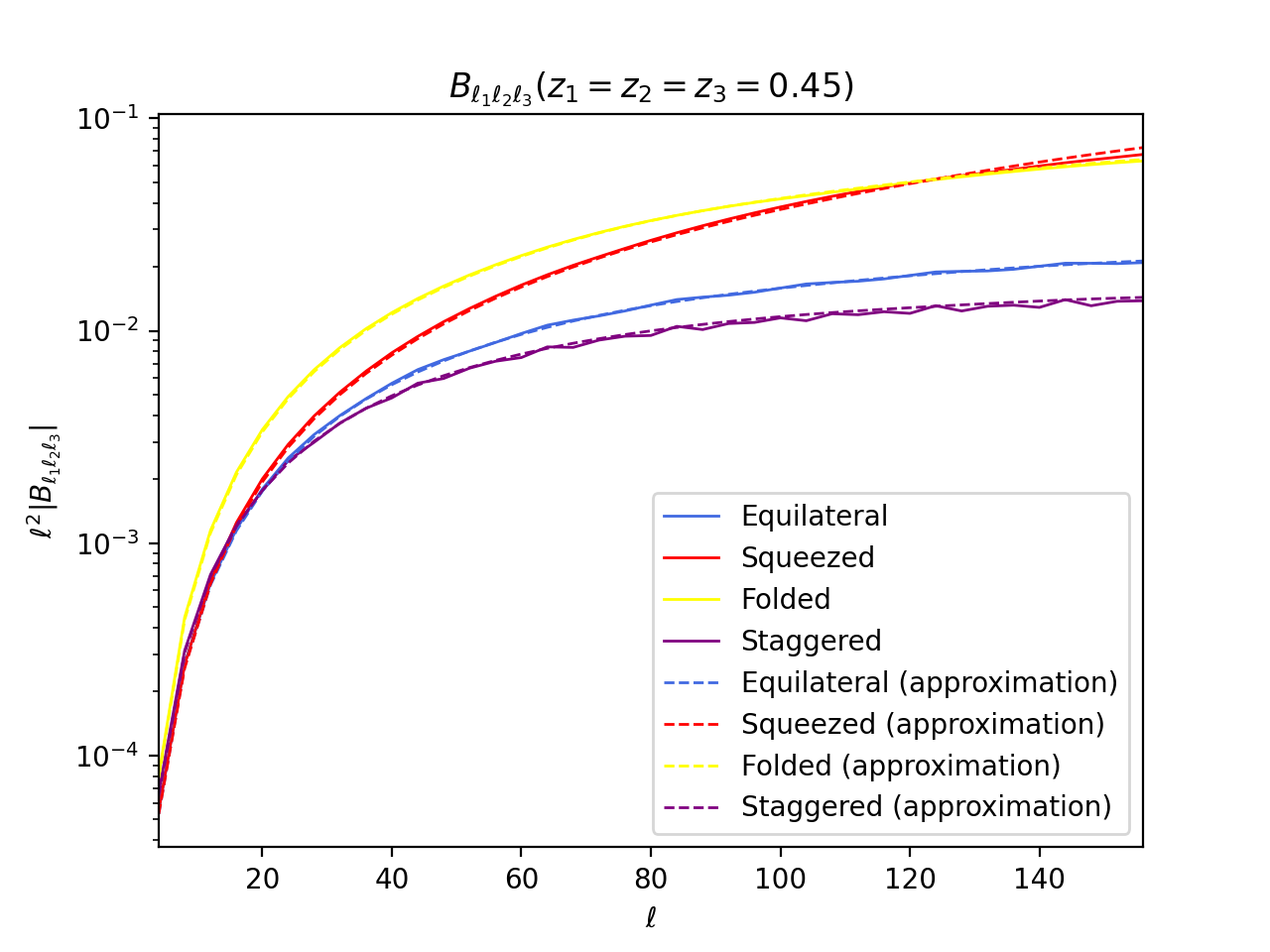}
		\label{fig:subfigb}
	\end{subfigure}
	\begin{subfigure}{0.32\linewidth}
		\includegraphics[width=\linewidth]{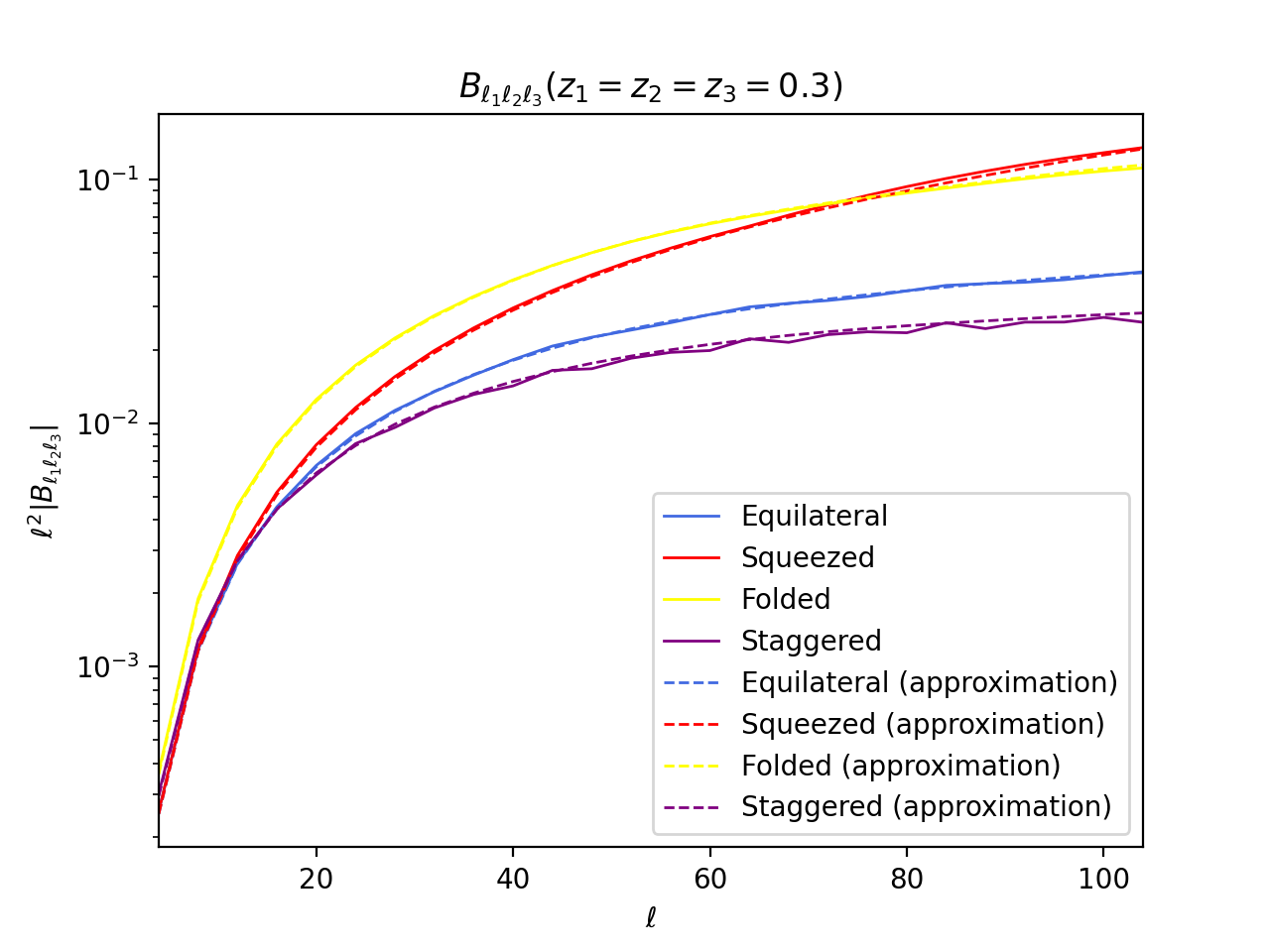}
		\label{fig:subfigc}
	\end{subfigure}
	\begin{subfigure}{0.32\linewidth}
	    \includegraphics[width=\linewidth]{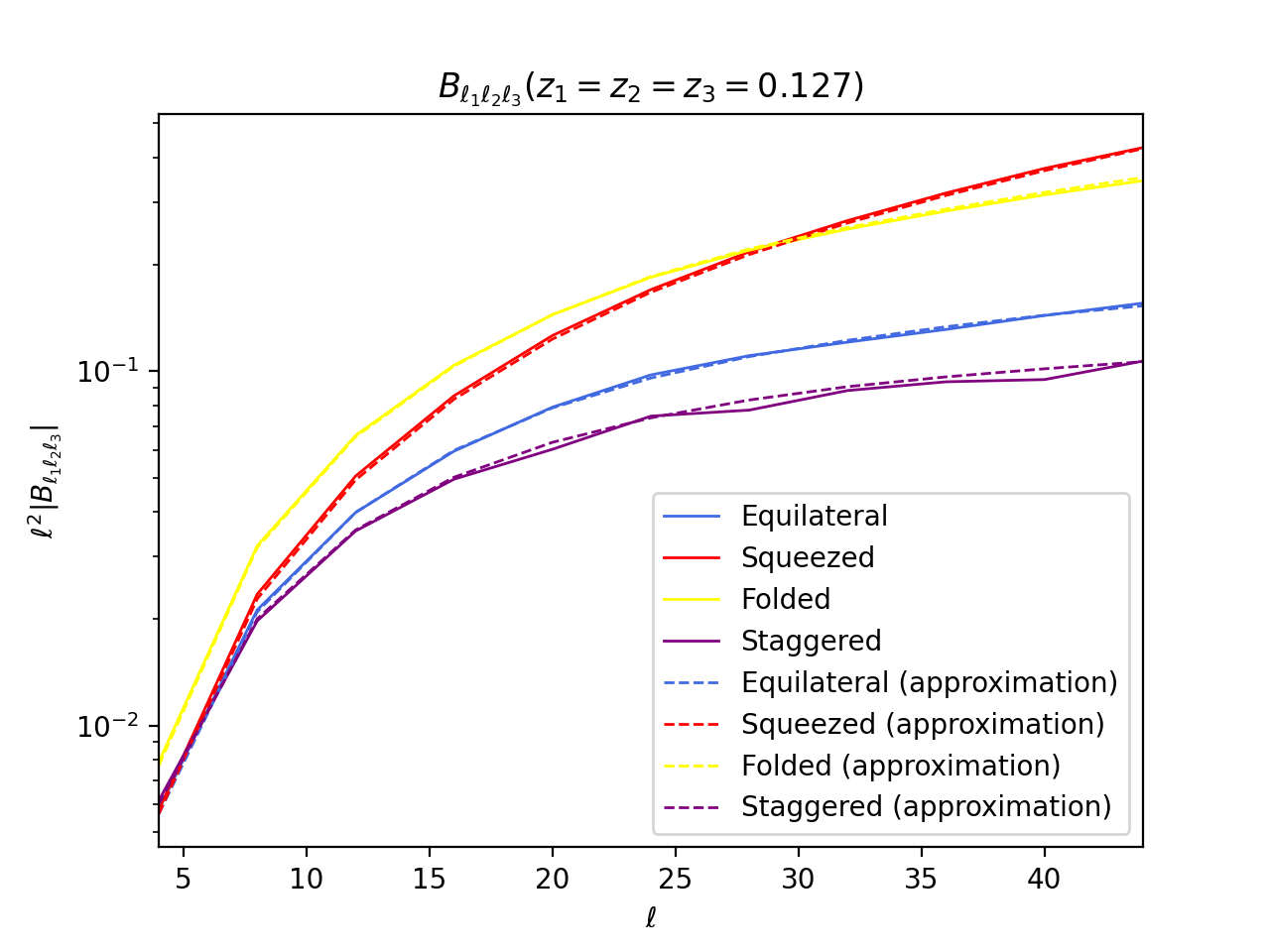}
	    \label{fig:subfigd}
    \end{subfigure}
    \begin{subfigure}{0.32\linewidth}
	    \includegraphics[width=\linewidth]{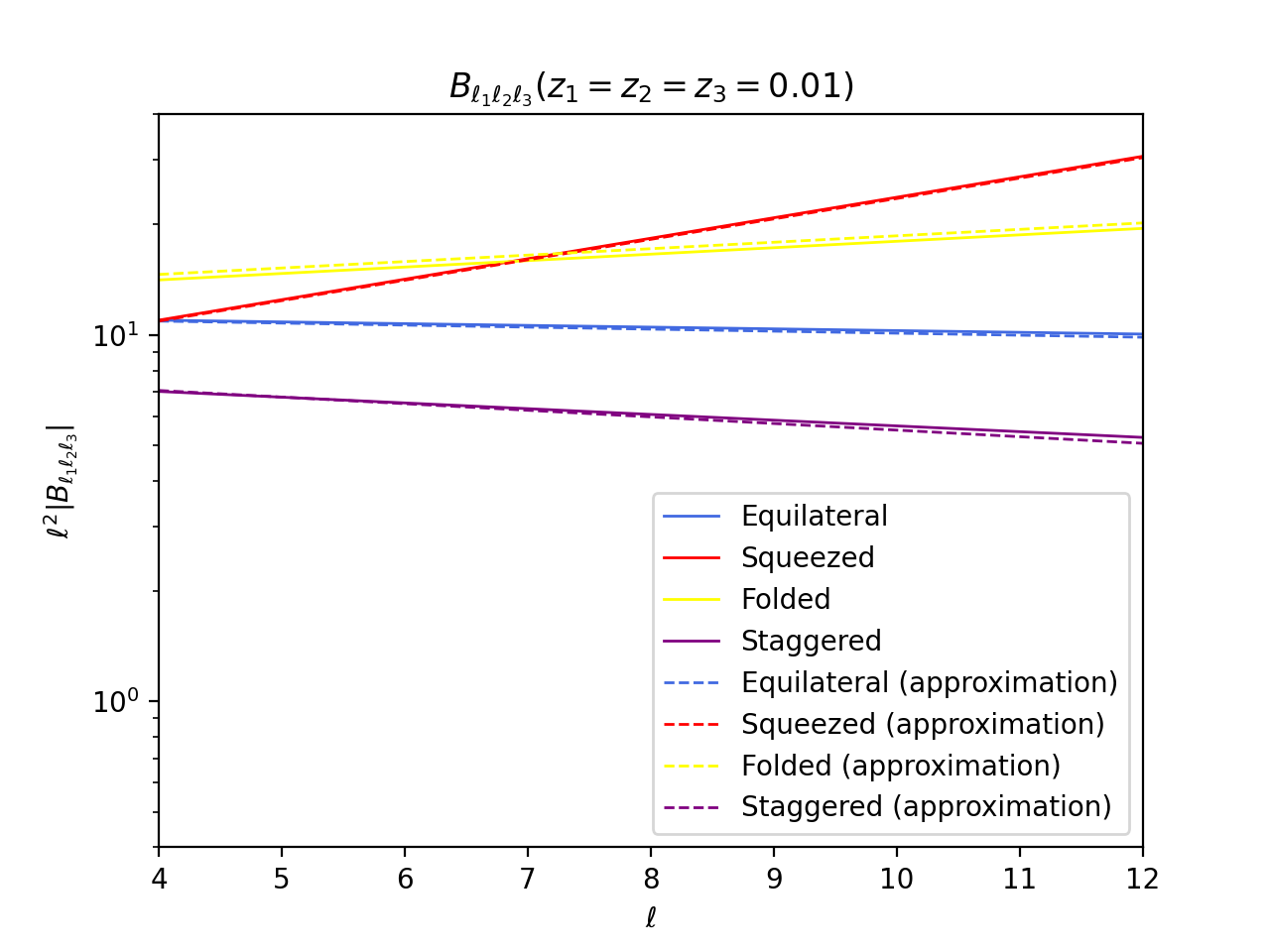}
	    \label{fig:subfige}
    \end{subfigure}

	\caption{Exact (solid) and approximate (dashed) total angle-averaged 21cm bispectrum at $z = 0.01, 0.127, 0.3, 0.45$, and $0.49$, shown for the equilateral, squeezed, folded, and staggered triangle configurations. The maximum multipole $\ell_{\rm max}$ is given by the nonlinear cutoff scale $\ell_{\rm max}^{\rm nl}$ at each redshift, except for $z=0.01$, where $\ell_{\rm max}$ is extrapolated to higher multipoles for better visualization. The approximate bispectrum is obtained using the averaged fractional parameter $r = 0.24$.}
	\label{fig:estimation}
\end{figure*}

Adopting a constant value $r=0.24$ at all redshifts leads to negligible changes in the accuracy of the approximation, with the maximum mean error and variance remaining the same as the previous values, $2.2\%$ and $1.8\%$, respectively; these values are reported in parentheses in Table~\ref{table:approximation}. In Fig.~\ref{fig:estimation}, we compare the exact and reconstructed bispectra obtained using $r=0.24$, for the same redshifts and triangle configurations used in Table~\ref{table:approximation}. The figure provides a graphical representation of the table results and shows that the approximation closely reproduces the exact bispectrum, with differences remaining small over the multipole range relevant for our analysis.

In the Fisher matrix computation, we adopt this approximation in order to significantly reduce the numerical cost of the bispectrum evaluation.

\section{HI intensity mapping}
\label{sec:survey}

\subsection{Intensity mapping surveys}

In order to forecast the power spectrum and the ability of the bispectrum to constrain cosmological parameters, we consider two next-generation single-dish intensity mapping surveys, BINGO \cite{Wuensche:2021dcx} and SKA1-MID Band 2 \cite{SKA:2018ckk}. Their configuration is summarized in Table~\ref{table:survey}.

The Baryon Acoustic Oscillations from Integrated Neutral Gas Observations (BINGO) radio telescope is a single-dish experiment located in Brazil, designed to map neutral hydrogen and to be among the first instruments to detect Baryon Acoustic Oscillations (BAO) at radio frequencies using the intensity mapping technique. The instrument has a system temperature $T_{{\rm sys}} = 70 \,{\rm K}$, an angular resolution of $ 40 \, {\rm armin}$, and a redshift coverage $0.13 < z < 0.45$, corresponding to the epoch in which dark energy dominates the cosmic expansion.

The first phase of the Square Kilometre Array (SKA1-MID) is a dish array located in South Africa aimed to probe the late-time Universe and the nature of dark energy. Although primarily an interferometric facility, it can also operate in single-dish autocorrelation mode for 21cm intensity mapping. The array of 15 m dishes enables sky coverage of several thousand square degrees over the redshift range $0.01  < z < 0.5$, corresponding to Band 2 frequencies, with a significantly lower system temperature than current single-dish experiments, providing high sensitivity to large-scale structure.

\begin{table*}

\centering
\caption{Fiducial parameters for BINGO and SKA1-MID telescopes.}
\begin{tabular}{lcc}
\hline
Parameters & BINGO & SKA1-MID Band 2\\ 
\hline                               
Frequency range   & [980, 1260] MHz & [950, 1410] MHz\\
Redshift range  &  [0.13, 0.45] & [0.01, 0.5]\\ 
Number of frequency channels &  28 & 46\\
Number of feed horns  &  28 & 197\\
Sky coverage with Galactic mask  &  2900 $\text{deg}^2$ & 5000 $\text{deg}^2$\\
Observational time ($t_{\text{obs}}$ )  &  1 year & 1 year\\
System temperature ($T_{\text{sys}}$ )  &  70 K & 15 K\\
Beam resolution ($\theta_{\text{FWHM}} $)  &  40 arcmin & 70 arcmin \\
\hline
\end{tabular}

\label{table:survey}
\end{table*} 

\subsection{Intensity mapping noise}

The most important contribution to the 21cm noise power spectrum is the thermal noise. For the single-dish mode, assuming that there is no correlation between different frequencies, the noise covariance is given by \cite{Bull:2014rha}
\begin{equation}
    \mathcal{N}_{l}(z) = \frac{T_{\text{sys}^2}}{\Delta \nu t_{\text{obs}}}\left(\frac{4\pi f_{\text{sky}}}{2 n_{\text{f}}}\right)\frac{1}{b_{\ell}(z)^{2}} \,,
\end{equation}
where $T_{\text{sys}}$ is the total system temperature (antenna plus sky temperatures), $\Delta \nu$ is the frequency channel width, $t_{\text{obs}}$ is the total time of observation, $n_{\text{f}}$ is the number of feed horns, $f_{\text{sky}}$ is the fraction of the sky observed, and $b_{\ell}^{2}$ is a factor that reduces the signal because of the telescope maximum beam resolution, which is given by 
\begin{equation}
    b_{\ell}(z_{i}) = \exp\left( -\frac{1}{2}\frac{\ell^{2} \theta^{2}_{\text{FWHM}}(\nu_{\text{center}})}{8 \ln 2}\frac{\nu^{2}_{\text{center}}}{\nu^{2}_{i}}\right)\,,
\end{equation}
where $\theta_{\text{FWHM}}$ is the beam resolution and $\nu_{\text{center}}$ is the middle frequency of the survey.

\subsection{The angular scale limits}
\label{subsec:angular-scale-limits}
The minimum scale a telescope can survey depends on the telescope maximum angular resolution and the modeling of nonlinearity. For single-dish radio telescopes like SKA and BINGO, the maximum multipole resolution is given by \cite{Bull:2014rha, Durrer:2020orn}:
\begin{equation}
    \ell^{\text{res}}_{\text{max}}(z) \approx \frac{2 \pi D_{d}}{\lambda_{21}(1+z)} \,,
\end{equation}
where $D_{d}$ is the dish diameter (40 m for BINGO, 15 m for SKA1-MID), and $\lambda_{21}$ is the rest-frame 21cm wavelength.

For low redshifts, there is a scale at which linear perturbation theory begins to break down. This nonlinear effect of matter clustering imposes a maximum multipole condition where we can perform our analysis using the power spectrum and bispectrum \cite{Jolicoeur:2020eup}:
\begin{align}
    \nonumber \ell^{\rm nl}_{\rm max}(z) &= r(z)k_{\rm nl}(z) \,, \\
    k_{\rm nl}(z) &=  k_{\rm nl}(0)(1 + z)^{2/3}  \,.
\end{align}
In our analysis we assume this maximum scale to be $k_{\rm nl}(0) = 0.2h/\text{Mpc}$, for the power spectrum, and $k_{\rm nl}(0) = 0.1h/\text{Mpc}$, for the bispectrum, where nonlinear effects are strongest.

\begin{figure}
    \centering
    \includegraphics[width=0.5\textwidth]{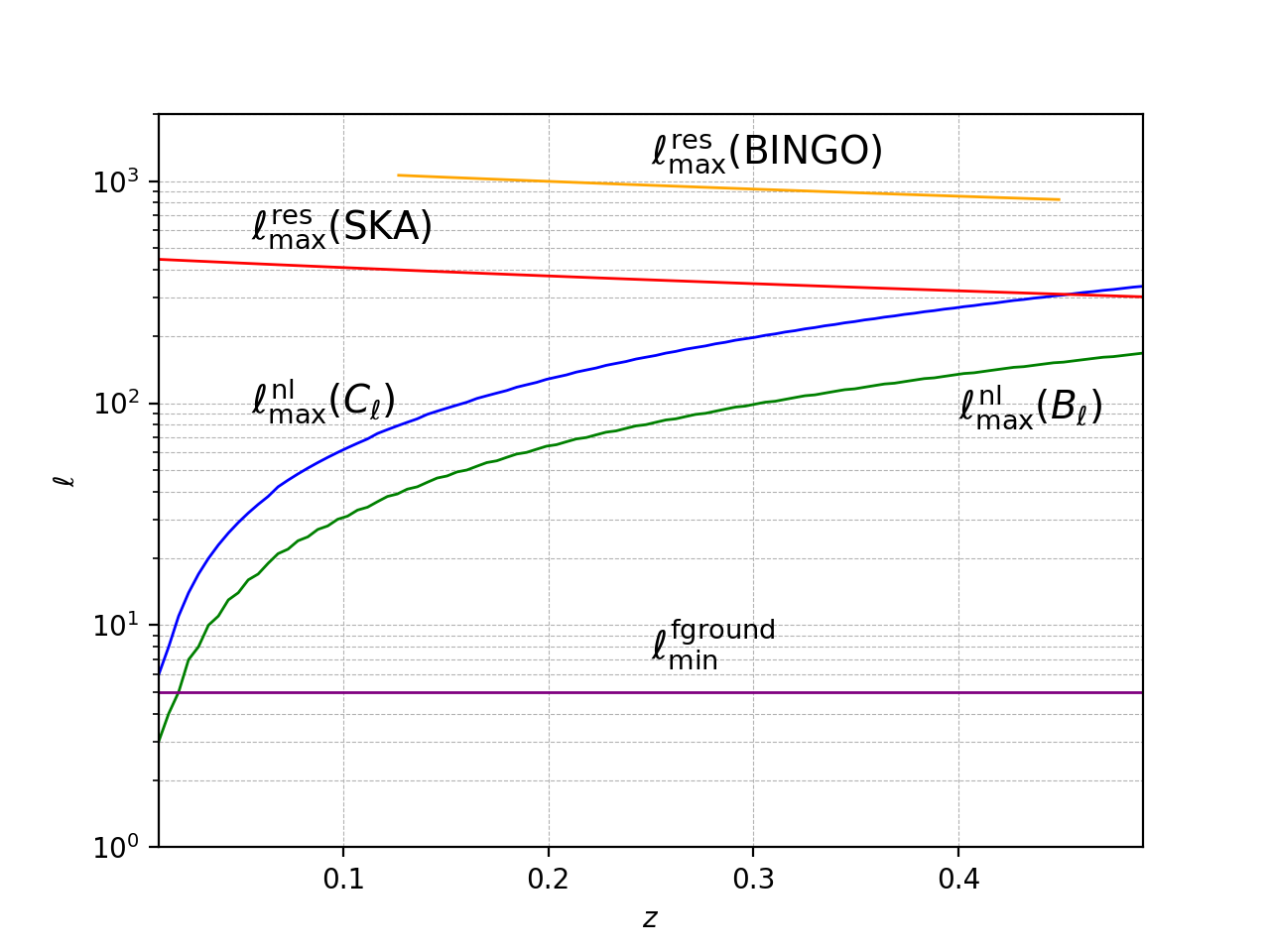}
    \caption{Specification of the multipole ranges [$\ell_{\text{min}}$, $\ell_{\text{max}}$] adopted for the SKA1-MID Band 2 and BINGO surveys. The lower bound, $\ell_{\text{min}}$, accounts for the loss of large-scale radial modes during foreground subtraction. We define $\ell_{\text{max}}$ to be determined by either the onset of the non-linear regime or the instrumental resolution.}
    \label{fig:l_min_max}
\end{figure}

Figure~\ref{fig:l_min_max} shows these multipole upward limits. In the BINGO and SKA redshift regime, the nonlinear effects are more important than the maximum angular resolution to set the survey minimum scale. The exception is for the SKA power spectrum in high redshifts. Therefore, in the calculation of Fisher forecast, we use the multipoles in range of $\ell^{\rm fground}_{\rm min}< \ell <\ell^{\rm nl}_{\rm max}(B_\ell)$ \footnote{In this work, $B_\ell$ is a shorthand notation for $B_{\ell_{1}\ell_{2}\ell_{3}}$.} for bispectrum and $\ell^{\rm fground}_{\rm min}< \ell <\ell^{\rm nl}_{\rm max}(C_\ell)$ for power spectrum. We choose a cutoff of the low mulitpoles $\ell^{\rm fground}_{\rm min} = 6$ that accounts for the loss of large-scale radial modes during foreground subtraction.

While we define formal $\ell_{\rm max}$ boundaries based on resolution or non-linearity, it was shown that thermal noise in the BINGO experiment increases sharply for $\ell>200$ \cite{Costa:2021jsk}; consequently, for high redshift, the effective information cutoff in the Fisher matrix analysis can be determined by the significant thermal noise rather than the formal $\ell_{\text{max}}$ boundaries of BINGO or SKA radio telescope.
 
\section{Fisher Matrix Formalism}
\label{sec:fisher}

In this section, we use the Fisher matrix formalism \cite{Fisher:1935abc, Tegmark:1996bz, Hobson:2009bmc} to forecast the capability of BINGO and SKA telescopes  to constrain the cosmological parameters of three models: $\Lambda$CDM, ${\rm w}$CDM and the Chevallier–Polarski–Linder (CPL) parameterization model using the 21cm angular power spectrum and bispectrum.

Table \ref{table:fiducial} summarizes the fiducial values and physical interpretations of the parameters considered in our Fisher analysis. For the primary cosmological parameters, we adopt the best-fit values from the Planck 2018 results \cite{Planck:2018vyg}. The HI sector is modeled using the energy density measured by \cite{Switzer:2013ewa} and a non-linear bias scheme defined in Eqs.~\eqref{bs}--\eqref{b2}. This bias model utilizes three constant amplitudes ($\bar{b}_1,\bar{b}_2,\bar{b}_s$), allowing for overall amplitude freedom while preserving the characteristic redshift evolution of the HI bias. For notational brevity, we hereafter omit the overbars from these parameters ($\bar{b}_1,\bar{b}_2,\bar{b}_s \rightarrow b_1,b_2,b_s$); all subsequent mentions refer to the effective HI bias amplitudes, for which we assume a fiducial value of unity.

\begin{table*}
\caption{Fiducial values and descriptions of the cosmological and astrophysical parameters used in the Fisher matrix analysis. The parameters $b_1,b_2,b_s$, and $\Omega_{{\rm HI}}$ represent the non-linear bias and the HI energy density, accounting for the 21cm astrophysics and non-linear clustering.}

\begin{tabular}{clc}\toprule

Parameter           & Physical Meaning                                      & Fiducial value \\ 
\midrule                             
$\Omega_{b}h^{2}$   & Baryon density at $z = 0$                    & 0.022383 \\
$\Omega_{c}h^{2}$   & Cold dark matter density at $z = 0$          & 0.12011 \\
$\ln(10^{10}A_{s})$ & Natural log of the scalar amplitude          & 3.044\\
$n_{s}$             & Spectral tilt                                & 0.96605 \\
$h$                 & Hubble parameter divide by 100               & 0.6732  \\
${\rm w}_{0}$        & Dark energy equation of state parameter      & -1 \\
${\rm w}_{a}$        & Dark energy evolution parameter                   & 0 \\
$b_{1}$             & Linear amplitude of the HI bias parameter    & 1.0 \\
$b_{2}$             & Nonlinear amplitude of the HI bias parameter & 1.0 \\
$b_{s}$             & Amplitude of the HI tidal bias parameter     & 1.0 \\ 
$\Omega_{{\rm HI}}$       & $\text{H}_{\text{I}}$ density at $z = 0$                        & 6.2 $\times 10^{-4}$ \\ \bottomrule

\end{tabular}
\label{table:fiducial}
\end{table*} 

The Fisher matrix is given by the ensemble average of the second derivative of the log-likelihood with respect to the parameters $p_{i}$ and $p_{j}$, 
\begin{equation}
    F_{ij}  \equiv - \left \langle \frac{\partial^{2} \ln{L(\bf{D},\bf{p})}}{\partial p_{i} \partial p_{j}} \right \rangle, 
\end{equation}
where $\bf{D}$ is the data vector, which can be $\bf{C}$ for the power spectrum or $\bf{B}$ for the bispectrum. The inverse of the Fisher matrix gives us the covariance among the parameters and the square root of the diagonal elements are the $1\sigma$ marginalized constraints.

\subsubsection{Power Spectrum}

The Fisher matrix for the angular power spectrum is
\begin{equation}
    F_{ij} = \frac{1}{2} \textbf{Tr} \left[  \mathbf{C}^{-1}\frac{\partial \mathbf{C}}{\partial p_{i}} \mathbf{C}^{-1} \frac{\partial \mathbf{C}}{\partial p_{j}} \right] \,,
\end{equation}
where the covariance matrix $\mathbf{C}$ is the sum of signal and noise:
\begin{equation}
    \mathbf{C} = C_{\ell}(z_{i},z_{j}) + \mathcal{N}_{\ell}(z_{i},z_{j}) \,.
    \label{eq:C}
\end{equation}
For comparison with the bispectrum case, we will consider just redshift autocorrelations. The bispectrum is a three point correlation function, so the number of triangle configurations increases fast with the number of redshifts and multipoles, slowing down the bispectrum computations. To mitigate this problem we limit our analysis to equal redshift correlations. A more complete work for the BINGO and SKA angular power spectrum in combination with the Planck 2018 results can be seen in \cite{Chen:2019jms,Costa:2021jsk}.  

\subsubsection{Bispectrum}

Considering only autocorrelations among redshift bins, the presence of rotational and parity symmetries, the Fisher matrix for angular bispectra can be written as \cite{Babich:2004yc,Yadav:2007rk,Schmit:2018rtf}
\begin{multline}
    F_{ij} = f_{{\rm sky}}\sum_{m}\sum_{\ell_{3} \geq \ell_{2} \geq \ell_{1}} \frac{1}{\Delta_{\ell_{1}\ell_{2}\ell_{3}}} \sigma^{-2}_{\ell_{1}\ell_{2}\ell_{3}}(z_{m},z_{m},z_{m}) \\
    \times \frac{\partial }{\partial p_{i}}B_{\ell_{1}\ell_{2}\ell_{3}}(z_{m},z_{m},z_{m}) \frac{\partial }{\partial p_{j}}B_{\ell_{1}\ell_{2}\ell_{3}}(z_{m},z_{m},z_{m}),
\end{multline}
where $\Delta_{\ell_{1}\ell_{2}\ell_{3}} $ are the symmetry factors that can be 1 if $\ell_{1} \neq \ell_{2} \neq \ell_{3}$, 6 if $\ell_{1} = \ell_{2} = \ell_{3}$, and 2 otherwise. $\sigma^{2}_{\ell_{1}\ell_{2}\ell_{3}}(z_{m},z_{m},z_{m})$ is the bispectrum covariance for equal redshifts, it combines signal and noise through products of power spectrum covariances as follows:    
\begin{equation}
\sigma^{2}_{\ell_{1},\ell_{2},\ell_{3}}(z_{m},z_{m},z_{m}) = \tilde{C}^{m m}_{\ell_{1}}\tilde{C}^{m m}_{\ell_{2}}\tilde{C}^{m m}_{\ell_{3}} \,,
\end{equation}
where $\tilde{C}^{ij}_{\ell}$ are the components of the covariance matrix, Eq.~\eqref{eq:C}. 

To speed up the calculations, we sampled the bispectrum with a linear multipole bin-width $\Delta_{l} = 10$. The final Fisher matrix is rescaled by a proportionality factor, and we can write
\begin{equation}
    F_{i j} \approx \frac{n_{{\rm total}}}{n_{{\rm sample}}} F_{i j, {\rm sample}},
\end{equation}
where $n_{{\rm total}}$ is the total number of possible non-vanishing configurations, $n_{{\rm sample}}$ is the number of sampled configurations drawn from this set, and $F_{i j, {\rm sample}}$ is the Fisher matrix evaluated using the sampled subset. 

{\it Numerical implementation.} In this work, we developed our own code to compute the power spectrum, the relativistic bispectrum and their associated Fisher information matrix. The pipeline was validated against established benchmarks in the literature, ensuring consistency with previous theoretical results \cite{Durrer:2020orn, DiDio:2018unb}. A primary advantage of this code is its ability to compute the bispectrum integrated over redshift bins, a prerequisite for direct comparison with observational data. To optimize performance for cosmological parameter estimation, the Fisher matrix calculation is fully parallelized across both multipoles and redshifts, significantly reducing the computational overhead.

\section{Results}
\label{sec:results}

Table~\ref{table:BINGO} presents the results for $\Lambda$CDM, ${\rm w}$CDM, and the CPL parameterization model using the BINGO configuration, while Table~\ref{table:SKA} shows the corresponding constraints obtained with the SKA1-MID Band 2 configuration. Both tables contain the forecasts of marginalized 1$\sigma$ errors on the cosmological parameters derived from the Fisher matrix, considering the Planck 2018 data alone, and the following combination of the 21cm statistics with Planck: $C_{\ell}$ + Planck, $B_{\ell}$ + Planck, and $C_{\ell}$ + $B_{\ell}$ + Planck. 

In general, for each column, the first value corresponds to the 1$\sigma$ constraint, the value in parenthesis denotes the relative percentage error, defined as $100\% \times \sigma/p_{i}^{\rm fid}$, where $p_{i}^{\rm fid}$ is the fiducial value of the parameter, and the value in square bracket indicates the improvement of the joint analysis relative to Planck-only constraints. 

To highlight the impact of including the bispectrum in addition to the $C_{\ell}$ + Planck  Fisher matrix, the last column reports the improvement obtained when using $C_{\ell} + B_{\ell}$ + Planck with respect to $C_{\ell}$ + Planck. This improvement is defined as $100 \% \times |\sigma_{C_{\ell} + B_{\ell} + {\rm Planck}} - \sigma_{C_{\ell} + {\rm Planck}}|/\sigma_{C_{\ell} + {\rm Planck}}  $. 

\subsection{BINGO forecast}

In the context of the BINGO Telescope, the angular power spectrum within the $\Lambda$CDM framework contributes to improving the constraints on the parameters $\Omega_{c}h^2$, $h$, $\Omega_{b}h^{2}$, and $n_s$ relative to those obtained from Planck data alone, as shown in Table~\ref{table:BINGO}. The two-point statistics provide a moderate improvement for $\Omega_{c}h^2$ (13$\%$) and $h$ (15$\%$), and a smaller contribution for $\Omega_{b}h^{2}$ (6.7$\%$) and $n_s$ (4.6$\%$). When the parameter space is extended, the impact shifts toward late-time cosmological parameters. In the ${\rm w}$CDM model, the improvement is dominated by $h$ (87$\%$) and ${\rm w}$ (82$\%$), while in the CPL parametrization the main gains are found for $h$ (68$\%$), ${\rm w}_0$ (17$\%$), and ${\rm w}_a$ (12$\%$). These results indicate that the 21cm angular power spectrum is particularly sensitive to the recent expansion history, providing significant information on the Hubble parameter and the dynamics of dark energy.

Overall, the combinations $C_\ell$ + Planck and $B_{\ell}$ + Planck yield comparable constraints for the parameters $\Omega_{b}h^{2}$, $\Omega_{c}h^2$, $A_s$, and $n_s$ across all considered models. In the $\Lambda$CDM scenario, the Hubble parameter is also similarly constrained by both combinations. The linear bias $b_1$ is better determined by $C_\ell$ + Planck in $\Lambda$CDM, with an uncertainty approximately a factor of two smaller than that obtained with $B_\ell$ + Planck. In the ${\rm w}$CDM model, the equation-of-state parameter ${\rm w}$ is constrained at a similar level (about twice as tight), while the parameters $h$ and $b_1$ are constrained of roughly a factor of three. Conversely, in the CPL framework, the bispectrum provides significantly tighter constraints on the dynamical dark energy parameters ${\rm w}_0$ and ${\rm w}_a$ compared to the power spectrum.

When the bispectrum is added to the $C_\ell$ + Planck combination, modest improvements are observed in $\Lambda$CDM for $\Omega_{c}h^2$ (7.1$\%$), and $h$ (5.3$\%$). In the ${\rm w}$CDM model, these parameters show improvement of 7.1$\%$ and 8.3$\%$, respectively, while a small additional gain is also found for the linear bias parameter (5$\%$). The most significant impact appears in the CPL parametrization, where the joint inclusion of the bispectrum leads to noticeably tighter constraints on the dark energy parameters ${\rm w}_0$ (74$\%$) and ${\rm w}_a$ (73$\%$), as well as on the Hubble parameter (57$\%$), with a slight improvement in the linear bias parameter (7.1$\%$). These findings demonstrate that the angular bispectrum plays a key role for BINGO in probing the nature of dark energy and the cosmic expansion history. 

Parameters associated with early-universe physics, such as $A_s$, $n_s$ and $\Omega_{b}h^2$, are already tightly constrained by CMB measurements from Planck across all considered models. Within the $\Lambda$CDM framework, however, these parameter exhibit small improvement ($<7\%$) when the angular power spectrum and bispectrum are included. This indicates that 21cm intensity mapping retains some sensitivity to early-universe physics, even though the dominant constraints remain CMB-driven.

\begin{table*}
\centering
\caption{Expected $1 \sigma$ constraints on the $\Lambda$CDM, ${\rm w}$CDM and CPL cosmological parameters from Planck, $C_{\ell}$ $+$ Planck, $B_{\ell}$ $+$ Planck, and $C_{\ell} + B_{\ell}$ + Planck, in the BINGO context. The result in parenthesis is the percent relative constraint given by $100 \% \times \sigma/p^{\text{fid}}_{i}$, and the result in the square bracket is the improvement of the joint analyses with respect to Planck only. The last column shows the improvement in combining $C_{\ell} + B_{\ell}$ + Planck with respect to $C_{\ell}$ $+$ Planck, it is  given by $100 \% \left| \sigma_{C_{\ell} + B_{\ell} + {\rm Planck}} - \sigma_{C_{\ell} + {\rm Planck}}\right|/\sigma_{C_{\ell} + {\rm Planck}}$.}

\vspace{0.5cm}
\begin{tabular}{cllllc}\toprule

Parameter & Planck & $C_{\ell} + \text{Planck}$  & $B_{\ell} + \text{Planck}$ & $C_{\ell} + B_{\ell} + \text{Planck}$  & Improvement \\

\addlinespace[0.5ex]

\hline
 \addlinespace[0.5ex]
 \multicolumn{6}{c}{$\Lambda$CDM }  \\
 \addlinespace[0.5ex]
\hline

\addlinespace[0.5ex]

$\Omega_{b}h^{2}$   &0.00015 (0.67$\%$) & 0.00014 (0.63$\%$) [6.7$\%$] & 0.00015 (0.67$\%$) [0$\%$]  & 0.00014 (0.63$\%$) [6.7$\%$] & 0$\%$\\
$\Omega_{c}h^{2}$   &0.0016 (1.3$\%$)   & 0.0014 (1.2$\%$) [13$\%$]    & 0.0015 (1.2$\%$) [6.3$\%$]  & 0.0013 (1.1$\%$) [19$\%$]    & 7.1$\%$\\
$\ln(10^{10}A_{s})$ &0.017 (0.56$\%$)   & 0.017 (0.56$\%$) [0$\%$]     & 0.017 (0.56$\%$) [0$\%$]    & 0.016 (0.53$\%$) [5.9$\%$]   & 5.9$\%$\\
$n_{s}$             &0.0044 (0.46$\%$)  & 0.0042 (0.43$\%$) [4.6$\%$]  & 0.0043 (0.45$\%$) [2.3$\%$] & 0.0041 (0.42$\%$) [6.8$\%$]  & 2.4$\%$\\
$h$                 &0.0067 (1$\%$)     & 0.0057 (0.85$\%$) [15$\%$]   & 0.0064 (0.95$\%$) [4.5$\%$] & 0.0054 (0.8$\%$) [19$\%$]    & 5.3$\%$\\ 
$b_{1}$             & -                 & 0.012 (1.2$\%$)              & 0.026 (2.6$\%$)             & 0.012 (1.2$\%$)              & 0$\%$\\ 
$b_{2}$             & -                 & -                            & 0.26 (26$\%$)               & 0.12  (12$\%$)               & - \\ 
$b_{s}$             & -                 & -                            & 0.32 (32$\%$)               & 0.30 (30$\%$)                & -  \\ 

\hline
 \addlinespace[0.5ex]
 \multicolumn{6}{c}{${\rm w}$CDM }  \\
\addlinespace[0.5ex]
\hline

\addlinespace[0.5ex]

$\Omega_{b}h^{2}$   &0.00015 (0.67$\%$) & 0.00015 (0.67$\%$) [0$\%$]  & 0.00015 (0.67$\%$) [0$\%$]  & 0.00015 (0.67$\%$) [0$\%$]  & 0 $\%$\\
$\Omega_{c}h^{2}$   &0.0014 (1.2$\%$)   & 0.0014 (1.2$\%$) [0$\%$]    & 0.0014 (1.2$\%$) [0$\%$]    & 0.0013 (1.1$\%$) [7.1$\%$]  & 7.1$\%$\\
${\rm w}$           &0.26 (26$\%$)      & 0.048 (4.8$\%$) [82$\%$]    & 0.11 (11$\%$) [58$\%$]      & 0.047 (4.7$\%$) [82$\%$]    & 2.1$\%$\\
$\ln(10^{10}A_{s})$ &0.016 (0.53$\%$)   & 0.016 (0.53$\%$) [0$\%$]    & 0.016 (0.53$\%$) [0$\%$]    & 0.016 (0.53$\%$) [0$\%$]    &  0 $\%$\\
$n_{s}$             &0.0044 (0.46$\%$)  & 0.0043 (0.45$\%$) [2.3$\%$] & 0.0043 (0.45$\%$) [2.3$\%$] & 0.0042 (0.43$\%$) [4.6$\%$] & 2.3$\%$\\
$h$                 &0.089 (13.22$\%$)  & 0.012 (1.8$\%$) [87$\%$]    & 0.035 (5.2$\%$) [61$\%$]    & 0.011 (1.6$\%$) [88$\%$]    & 8.3 $\%$\\ 
$b_{1}$             & -                 & 0.04 (4$\%$)                & 0.13 (13$\%$)               & 0.038 (3.8$\%$)             & 5$\%$\\ 
$b_{2}$             & -                 & -                           & 0.26 (26$\%$)               & 0.13  (13$\%$)              & - \\ 
$b_{s}$             & -                 & -                           & 0.94 (94$\%$)               & 0.41 (41$\%$)               & - \\

\addlinespace[0.5ex]

\hline
\addlinespace[0.5ex]
 \multicolumn{6}{c}{CPL}  \\
 \addlinespace[0.5ex]
\hline

\addlinespace[0.5ex]

$\Omega_{b}h^{2}$   &0.00016 (0.71$\%$) & 0.00015 (0.67$\%$) [6.3$\%$] & 0.00015 (0.67$\%$) [6.3$\%$] & 0.00015 (0.67$\%$) [6.3$\%$] & 0$\%$\\
$\Omega_{c}h^{2}$   &0.0013 (1.1$\%$)   & 0.0013 (1.1$\%$) [0$\%$]     & 0.0013 (1.1$\%$) [0$\%$]     & 0.0013 (1.1$\%$) [0$\%$]     & 0$\%$\\
${\rm w}_{0}$       &0.46 (46$\%$)      & 0.38 (38$\%$) [17$\%$]       & 0.13 (13$\%$) [72$\%$]       & 0.1 (10$\%$) [78$\%$]        & 74$\%$\\
${\rm w}_{a}$       &1.7                & 1.5 [12$\%$]                 & 0.46 [73$\%$]                & 0.4 [76$\%$]                 & 73$\%$\\
$\ln(10^{10}A_{s})$ &0.016 (0.53$\%$)   & 0.016 (0.53$\%$) [0$\%$]     & 0.016 (0.53$\%$) [0$\%$]     & 0.016 (0.53$\%$) [0$\%$]     & 0$\%$\\
$n_{s}$             &0.0044 (0.46$\%$)  & 0.0044 (0.46$\%$) [0$\%$]    & 0.0044 (0.46$\%$) [0$\%$]    & 0.0043 (0.45$\%$) [2.3$\%$]  & 2.3$\%$\\
$h$                 &0.088 (13$\%$ )    & 0.028 (4.2$\%$) [68$\%$]     & 0.037 (5.5$\%$) [58$\%$]     & 0.012 (1.8$\%$) [86$\%$]     & 57$\%$\\ 
$b_{1}$             & -                 & 0.042 (4.2$\%$)              & 0.16 (16$\%$)                & 0.039 (3.9$\%$)              & 7.1$\%$\\ 
$b_{2}$             & -                  & -                           & 0.4 (40$\%$)                 & 0.27  (27$\%$)               & - \\ 
$b_{s}$             & -                  & -                           & 0.98 (98$\%$)                & 0.49 (49$\%$)                & - \\ 

\addlinespace[0.5ex]

\bottomrule

\end{tabular}
\label{table:BINGO}
\end{table*}

\subsection{SKA1-MID Band 2 forecast}

The analysis for the SKA1-MID Band 2 configuration reveals a significantly enhanced constraining power compared to BINGO, due to its superior survey volume and higher signal-to-noise ratio. As shown in Table~\ref{table:SKA}, the SKA angular power spectrum within the $\Lambda$CDM framework improves constraints on $\Omega_{c}h^2$ and $h$ by roughly 30$\%$, doubling the relative improvement seen in the BINGO setup. In the ${\rm w}$CDM scenario, the improvement in $h$ (93$\%$) and ${\rm w}$ (85$\%$) is complemented by small but perceptible gains in $\Omega_{c}h^2$ (7.1$\%$) and $n_s$ (4.6$\%$). This trend holds in the CPL parametrization, where SKA provides robust information on $h$ (78$\%$) and substantial constraints on ${\rm w}_{0}$ and ${\rm w}_{a}$ ($\sim 44\%$), while maintaining sensitivity to the primordial spectral index (4.6$\%$).

The synergy between SKA statistics and Planck data follows a similar structural pattern to BINGO, yet with deeper precision. While the $C_{\ell}$ + Planck and $B_{\ell}$ + Planck statistics provide similar bounds for the core cosmological set $\{\Omega_{b}h^2,\Omega_{c}h^2,A_s,n_s\}$, the SKA bispectrum remains the superior probe for dynamical dark energy in the CPL framework.

The addition of the bispectrum Fisher matrix into the joint analysis, $C_{\ell}$ + Planck, yields the most significant results for the CPL model, where ${\rm w}_{0}$ and ${\rm w}_{a}$ see improvements of 77$\%$. Notably, the Hubble parameter constraint is refined by 63$\%$, and the parameters $\Omega_{c}h^2$, $\ln(10^{10}A_{s})$, $n_{s}$ and $b_1$ shows a healthy 5-7$\%$ improvement, reflecting the high-resolution capacity of the SKA.

Regarding the early Universe, SKA data allow for a more pronounced deviation from the Planck-only baseline. In contrast to BINGO, where the sensitivity to $A_{s}$, $n_s$ and $\Omega_{b}h^2$ was largely limited to the $\Lambda$CDM, SKA shows a cumulative improvement in the constraints on $n_s$ and $\Omega_{b}h^2$ across all three models. This improvement emerges when the power spectrum is first combined with Planck and becomes stronger when the full $C_{\ell}$ + $B_{\ell}$ + Planck Fisher matrix is used. In the $\Lambda$CDM case, the cumulative improvement is moderate, about 15-20$\%$, whereas for the extended models the improvement is smaller, at the level of 6-9$\%$. The parameter $\ln(10^{10} A_s)$ shows a different behavior across the models: in the $\Lambda$CDM model, the combination $C_{\ell}$ + Planck is already sufficient to reach the maximal improvement of about 6$\%$; for the wCDM model, as in the BINGO case, no improvement is observed. In the CPL model there is no improvement when using either $C_{\ell}$ + Planck or $B_{\ell}$ + Planck separately, but the combination of them allow for a 6.3$\%$ improvement. These results indicate that the large survey volume of SKA enables 21cm intensity mapping to probe early-Universe physics and to provide a meaningful refinement of CMB-derived constraints.

\begin{table*}

\centering
\caption{Expected $1 \sigma$ constraints on the $\Lambda$CDM, ${\rm w}$CDM and CPL cosmological parameters from Planck, $C_{\ell}$ $+$ Planck, $B_{\ell}$ $+$ Planck, and $C_{\ell} + B_{\ell}$ + Planck, in the SKA1-MID band-2 context. The result in parenthesis is the percent relative constraint given by $100 \% \times \sigma/p^{\text{fid}}_{i}$, and the result in the square bracket is the improvement of the joint analyses with respect to Planck only. The last column shows the improvement in combining $C_{\ell} + B_{\ell}$ + Planck with respect to $C_{\ell}$ $+$ Planck, it is  given by $100 \% \left| \sigma_{C_{\ell} + B_{\ell} + {\rm Planck}} - \sigma_{C_{\ell} + {\rm Planck}}\right|/\sigma_{C_{\ell} + {\rm Planck}}$. }

\vspace{0.5cm}
\begin{tabular}{cllllc}\toprule

Parameter & Planck & $C_{\ell} + \text{Planck}$&$B_{\ell} + \text{Planck}$& $C_{\ell} + B_{\ell} + \text{Planck}$ & Improvement\\

\addlinespace[0.5ex]

\hline
\addlinespace[0.5ex]
 \multicolumn{6}{c}{$\Lambda$CDM }  \\
\addlinespace[0.5ex]
\hline

\addlinespace[0.5ex]

$\Omega_{b}h^{2}$   &0.00015 (0.67$\%$) & 0.00013 (0.58$\%$) [13$\%$] & 0.00014 (0.63$\%$) [6.7$\%$] & 0.00012 (0.54$\%$) [20$\%$] & 7.7$\%$\\
$\Omega_{c}h^{2}$   &0.0016 (1.3$\%$)   & 0.0011 (0.92$\%$) [31$\%$]  & 0.0013 (1.1$\%$) [19$\%$]    & 0.00094 (0.78$\%$) [41$\%$] & 15$\%$\\
$\ln(10^{10}A_{s})$ &0.017 (0.56$\%$)   & 0.016 (0.53$\%$) [5.9$\%$]  & 0.016 (0.53$\%$) [5.9$\%$]   & 0.016 (0.53$\%$) [5.9$\%$]  & 0$\%$\\
$n_{s}$             &0.0044 (0.46$\%$)  & 0.0038 (0.39$\%$) [14$\%$]  & 0.0041 (0.42$\%$) [6.8$\%$]  & 0.0037 (0.38$\%$) [16$\%$]  & 2.6$\%$\\
$h$                 &0.0067 (1$\%$)     & 0.0044 (0.65$\%$) [34$\%$]  & 0.0057 (0.85$\%$) [15$\%$]   & 0.004 (0.59$\%$) [40$\%$]   & 9.1$\%$\\ 
$b_{1}$             & -                 & 0.01 (1$\%$)                & 0.019 (1.9$\%$)              & 0.01 (1$\%$)                & 0 $\%$\\ 
$b_{2}$             & -                 & -                           & 0.15 (15$\%$)                & 0.074 (7.4$\%$)             & - \\ 
$b_{s}$             & -                 & -                           & 0.20 (20$\%$)                & 0.19 (19$\%$)               & - \\ 

\addlinespace[0.5ex]

\hline
\addlinespace[0.5ex]
 \multicolumn{6}{c}{${\rm w}$CDM }  \\
\addlinespace[0.5ex]
\hline

\addlinespace[0.5ex]

$\Omega_{b}h^{2}$   &0.00015 (0.67$\%$) & 0.00015 (0.67$\%$) [0$\%$]   & 0.00014 (0.63$\%$) [6.7$\%$] & 0.00014 (0.63$\%$) [6.7$\%$] & 6.7$\%$\\
$\Omega_{c}h^{2}$   &0.0014 (1.2$\%$)   & 0.0013 (1.1$\%$) [7.1$\%$]   & 0.0013 (1.1$\%$) [7.1$\%$]   & 0.0012 (1$\%$) [14$\%$]      & 7.4$\%$\\
${\rm w}$           &0.26 (26$\%$)      & 0.039 (3.9$\%$) [85$\%$]     & 0.069 (6.9$\%$) [73$\%$]     & 0.037 (3.7$\%$) [86$\%$]     & 5.1$\%$\\
$\ln(10^{10}A_{s})$ &0.016 (0.53$\%$)   & 0.016 (0.53$\%$) [0$\%$]     & 0.016 (0.53$\%$) [0$\%$]     & 0.016 (0.53$\%$) [0$\%$]     & 0$\%$\\
$n_{s}$             &0.0044 (0.46$\%$)  & 0.0042 (0.43$\%$) [4.6$\%$]  & 0.0042 (0.43$\%$) [4.6$\%$]  & 0.004 (0.41$\%$) [9.1$\%$]   & 4.8$\%$\\
$h$                 &0.089 (13$\%$)     & 0.0065 (0.97$\%$) [93$\%$]   & 0.02 (3$\%$) [78$\%$]        & 0.0061 (0.91$\%$) [93$\%$]   & 6.2$\%$\\ 
$b_{1}$             & -                 & 0.026 (2.6$\%$)              & 0.077 (7.7$\%$)              & 0.025 (2.5$\%$)              & 3.8$\%$\\ 
$b_{2}$             & -                 & -                            & 0.15 (15$\%$)                & 0.075 (7.5$\%$)              & - \\ 
$b_{s}$             & -                 & -                            & 0.53 (53$\%$)                & 0.24 (24$\%$)                & - \\

\addlinespace[0.5ex]

\hline
\addlinespace[0.5ex]
 \multicolumn{6}{c}{CPL }  \\
\addlinespace[0.5ex]
\hline

\addlinespace[0.5ex]

$\Omega_{b}h^{2}$   &0.00016 (0.71$\%$) & 0.00015 (0.67$\%$) [6.3$\%$] & 0.00015 (0.67$\%$) [6.3$\%$] & 0.00015 (0.67$\%$) [6.3$\%$] & 0$\%$\\
$\Omega_{c}h^{2}$   &0.0013 (1.1$\%$)   & 0.0013 (1.1$\%$) [0$\%$]     & 0.0013 (1.1$\%$) [0$\%$]     & 0.0012 (1$\%$) [7.7$\%$]     & 7.4$\%$\\
${\rm w}_{0}$       &0.46 (46$\%$)      & 0.26 (26$\%$) [43$\%$]       & 0.08 (8$\%$) [83$\%$]        & 0.061 (6.1$\%$) [87$\%$]     & 77$\%$\\
${\rm w}_{a}$       &1.8                & 1 [44$\%$]                   & 0.28 [84$\%$]                & 0.23 [87$\%$]                & 77$\%$\\
$\ln(10^{10}A_{s})$ &0.016 (0.53$\%$)   & 0.016 (0.53$\%$) [0$\%$]     & 0.016 (0.53$\%$) [0$\%$]     & 0.015 (0.49$\%$) [6.3$\%$]   & 6.3$\%$\\
$n_{s}$             &0.0044 (0.46$\%$)  & 0.0042 (0.43$\%$) [4.6$\%$]  & 0.0043 (0.45$\%$) [2.3$\%$]  & 0.004 (0.41$\%$) [9.1$\%$]   & 4.8$\%$\\
$h$                 &0.088 (13$\%$ )    & 0.019 (2.8$\%$) [78$\%$]     & 0.021 (3.1$\%$) [76$\%$]     & 0.007 (1.04$\%$) [92$\%$]    & 63$\%$\\ 
$b_{1}$             & -                 & 0.027 (2.7$\%$)              & 0.095 (9.5$\%$)              & 0.025 (2.5$\%$)              & 7.4$\%$\\ 
$b_{2}$             & -                 & -                            & 0.23 (23$\%$)                & 0.15 (15$\%$)                & - \\ 
$b_{s}$             & -                 & -                            & 0.57 (57$\%$)                & 0.27 (27$\%$)                & - \\

\addlinespace[0.5ex]

\bottomrule

\end{tabular}
\label{table:SKA}
\end{table*} 


The significantly larger relative improvement observed when combining the angular power spectrum and bispectrum in the CPL model, compared to $\Lambda$CDM and ${\rm w}$CDM (see the last column of the Tables ~\ref{table:BINGO} and \ref{table:SKA}), can be understood as a consequence of the strong degeneracies introduced by a time varying dark energy equation of state. In the CPL parametrization, the parameters ${\rm w}_0$ and ${\rm w}_a$ primarily affect the redshift dependence of both the expansion history and the growth of structure, leading to elongated degeneracy directions in the power-spectrum Fisher matrix. 

\begin{figure*}[htbp]
	\centering
	\begin{subfigure}{0.48\linewidth}
		\includegraphics[width=\linewidth]{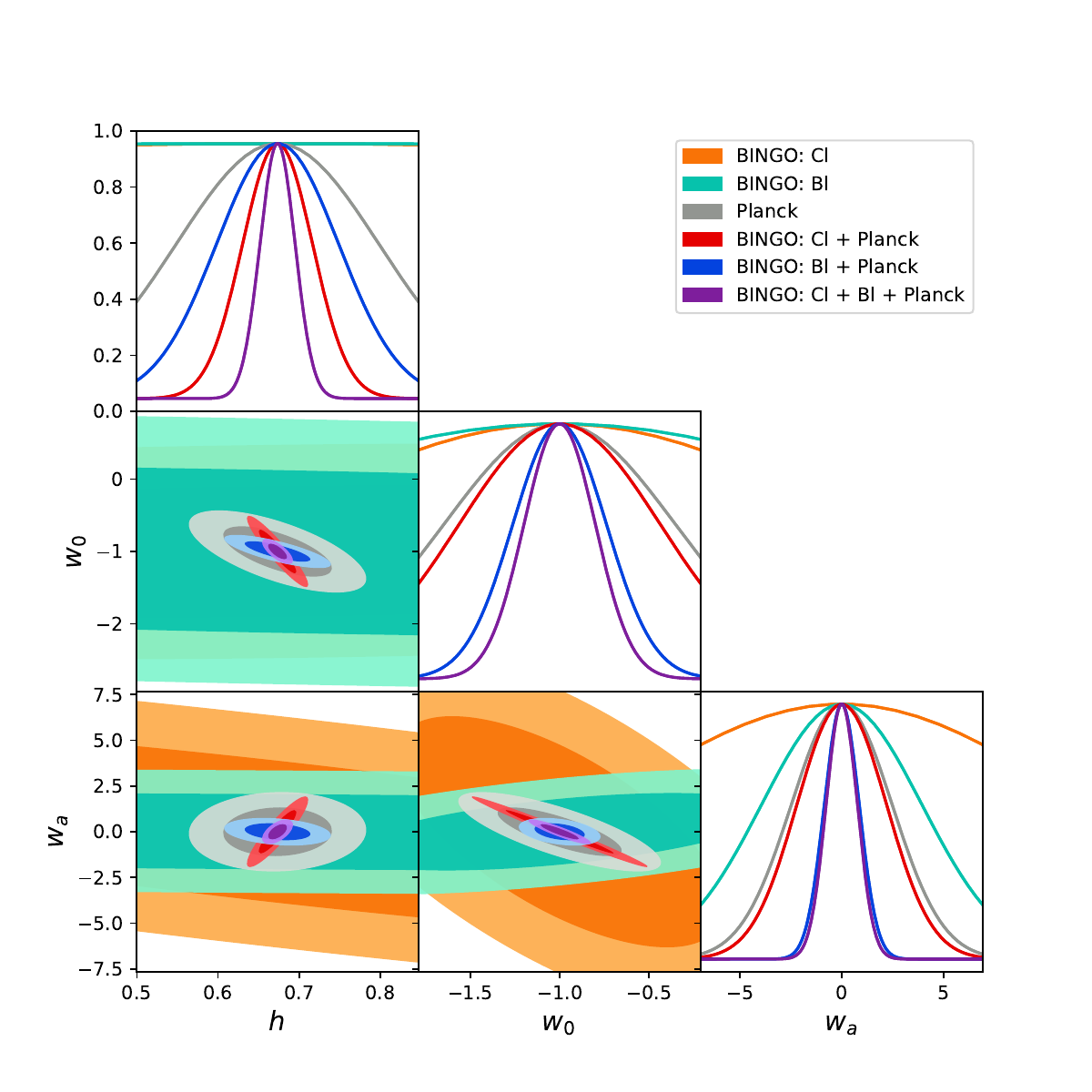}
		\label{fig:triangle_BINGO}
	\end{subfigure}
    \hfill
    \begin{subfigure}{0.48\linewidth}
		\includegraphics[width=\linewidth]{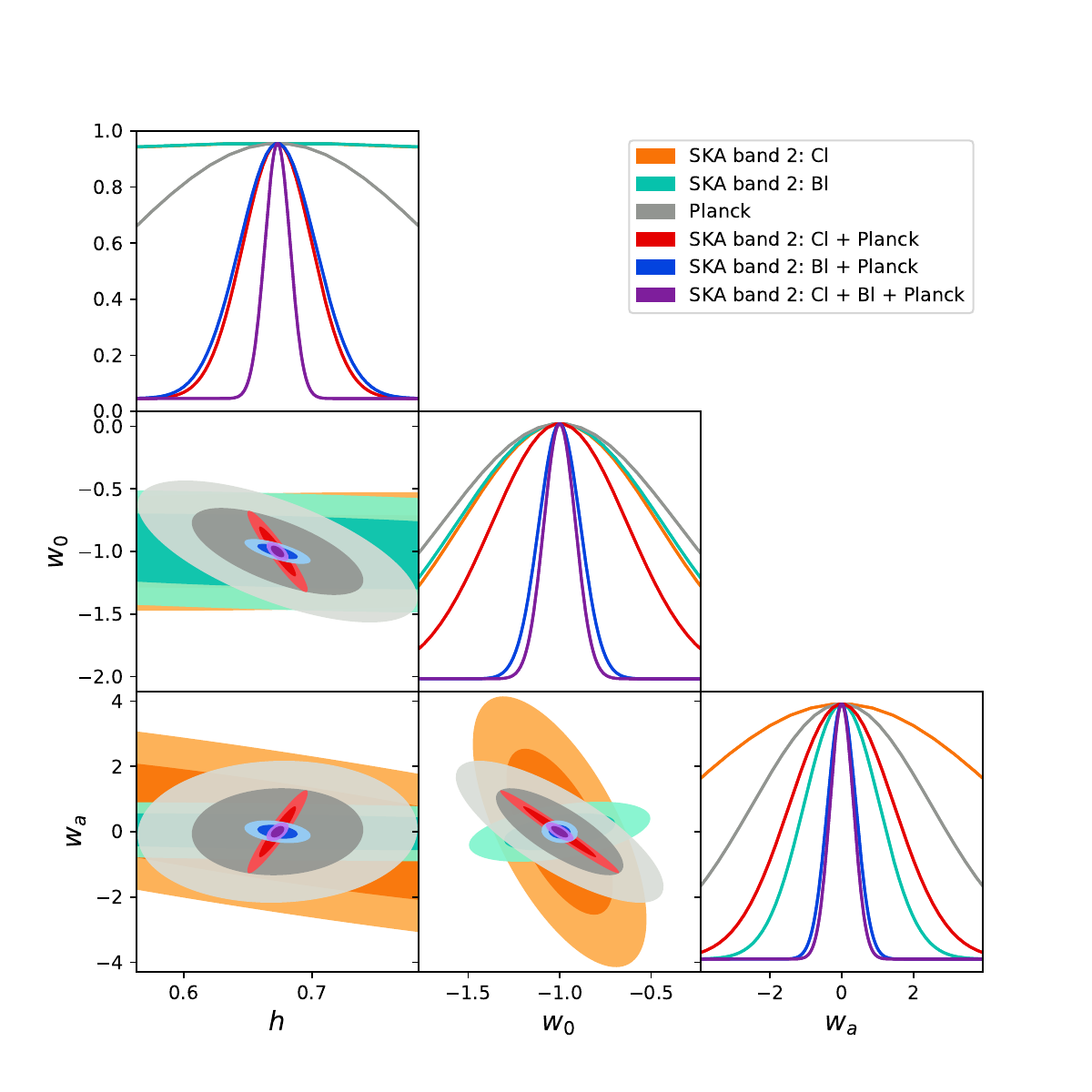}
		\label{fig:triangle_SKA}
	\end{subfigure}
	
	\caption{Forecasted 68$\%$ and 95$\%$ confidence contours for the CPL parameters ${\rm w}_0$, ${\rm w}_a$, and the Hubble constant $h$. The left panel corresponds to BINGO and the right panel to SKA1-MID band 2. Constraints are derived using the $C_\ell$, $B_{\ell}$, Planck, $C_\ell + \text{Planck}$, and $B_{\ell} + \text{Planck}$ statistics.}
	\label{fig:triangle}
\end{figure*}

The angular bispectrum probes nonlinear mode coupling and growth history with a parameter dependence that is significantly rotated relative to that of the angular power spectrum and Planck constraints, as can be seen in the Figure~\ref{fig:triangle} for BINGO and SKA cases. As a result, the joint $C_{\ell} + B_{\ell}$ analysis efficiently breaks degeneracies among ${\rm w}_0$, ${\rm w}_a$, and $h$, leading to a substantial reduction of the marginalized uncertainties. This effect is much weaker in $\Lambda$CDM and ${\rm w}$CDM, where the growth history is less flexible and the degeneracy directions of the power spectrum and bispectrum are more closely aligned. As this same behavior is observed in both the BINGO and the SKA radio telescope, this indicates that the improvement arises from degeneracy breaking rather than from an increase in signal-to-noise or redshift coverage.

\section{Conclusion}
\label{sec:conclusion}

This is the first time the relativistic bispectrum has been used to perform cosmological parameter estimation for the BINGO radio telescope, and, to the best of our knowledge, it is also the first time that the angular bispectrum without the Limber approximation has been applied to the SKA radio telescope. As shown in previous works, the second-order velocity component provides an important contribution to the bispectrum and is incompatible with the Limber approximation, except in cases where the redshift bins are significantly wide, in which case this component can be neglected at the cost of losing a substantial amount of cosmological information \cite{DiDio:2015bua, DiDio:2018unb}. In the present work, we show that the second-order velocity component accounts for 24$\%$ of the total signal at low redshifts.

To ensure the robustness of our forecasts for low-redshift experiments like BINGO and SKA1-MID Band 2, we implemented a conservative cut on non-linear scales. This approach represents a refinement over previous power spectrum analyses \cite{Chen:2019jms, Costa:2021jsk}, yielding more physically realistic constraints. We performed comprehensive forecasts for the $\Lambda$CDM and ${\rm w}$CDM models, as well as the Chevallier-Polarski-Linder (CPL) parametrization for dynamical dark energy.

Our findings demonstrate that the 21cm angular power spectrum significantly enhances constraints on expansion and dark energy parameters beyond Planck observations alone, with SKA1-MID Band 2 roughly doubling the constraining power of BINGO for $\Omega_c h^2$ and $h$ in the $\Lambda$CDM framework. While improvements for $\Omega_c h^2$, $\Omega_b h^2$, and $n_s$ are marginal or small across all models, both telescopes show a dramatic sensitivity to the recent expansion history, with $h$ and dark energy parameters seeing gains of up to $93\%$ in extended models. Our results align with those established in \cite{Costa:2021jsk}, though we report naturally broader constraints due to our exclusive use of redshift auto-correlations and the exclusion of non-linear scales.

For the $\Lambda$CDM and ${\rm w}$CDM models, the bispectrum (when combined with Planck) generally yields constraints comparable to those of the angular power spectrum. However, the power spectrum maintains superior sensitivity to the 21cm linear bias, the dark energy equation-of-state parameter ${\rm w}$, and the Hubble constant $h$. The most striking results emerge in the context of dynamical dark energy. For the CPL model, the bispectrum significantly outperforms two-point statistics in constraining ${\rm w}_0$ and ${\rm w}_a$. This underscores the necessity of accounting for relativistic non-Gaussianities when probing the nature of dark energy. Specifically, integrating the bispectrum Fisher matrix with the $C_{\ell}$ + Planck joint analysis results in:

- $\, >70 \%$ improvement in constraints for ${\rm w}_0$ and ${\rm w}_a$,

- \, $\approx 60\%$ improvement for the Hubble parameter $h$.

These gains are consistent across both BINGO and SKA1-MID Band 2 configurations, highlighting the bispectrum as a pivotal tool for next-generation radio cosmology.

The results presented in this work are particularly robust in the large-scale, weakly non-linear regime, where relativistic and projection effects are most relevant and where the full-sky treatment adopted here becomes essential. By avoiding the Limber approximation and restricting the analysis to scales where perturbation theory is valid, our pipeline provides a consistent description of the angular power spectrum and bispectrum, including second-order velocity terms and relativistic projection contributions that become important for thin redshift bins and require a full-sky treatment \cite{DiDio:2018unb}. This makes the present framework well suited for forecasts for intensity-mapping experiments such as BINGO and SKA, which are particularly sensitive to large angular scales.

Our analysis was intentionally conservative in the treatment of small-scale modes to ensure theoretical control. We imposed a non-linear cutoff and removed the lowest multipoles ($\ell \le 5$) to account for the loss of large-scale radial modes due to foreground cleaning. In practice, the transition to the non-linear regime is gradual, and higher-order corrections \cite{Baldauf:2016sjb} and non-Gaussian covariance terms may become relevant before the adopted cutoff \cite{Chan:2016ehg, Biagetti:2021tua}. Including these contributions would allow the use of additional modes, although the improvement in the constraints may be reduced once non-Gaussian covariance is properly taken into account.

We also restricted the forecast to redshift bins auto-correlations. This approximation significantly reduces the computational cost of the full-sky power spectrum and bispectrum calculation, though it does not capture the full radial information of the survey volume. Including cross-correlations among the redshift bins is expected to tighten the constraints, although at the price of a substantial increase in numerical complexity.

Another simplifying assumption concerns the modeling of the HI bias. We adopted redshift-dependent functional forms for $b_1(z)$, $b_2(z)$, and $b_s(z)$ with overall amplitudes treated as free parameters common to all redshift bins. This parametrization provides a simplified description of the bias uncertainty while keeping the parameter space manageable for a full-sky power spectrum and bispectrum Fisher analysis. Allowing the bias amplitudes to vary independently in each bin would provide a more realistic description of the signal, but would increase degeneracies and likely weaken the constraints.

A similar analysis targeting the SKA, HIRAX \cite{Crichton:2021hlc}, and MeerKAT \cite{MeerKLASS:2017vgf} radio telescopes was conducted by \cite{Karagiannis:2022ylq}. Their methodology employs a Fourier-space approach within the flat-sky (Limber) approximation, incorporating a non-linear cutoff, non-Gaussian corrections to the diagonal bispectrum covariance, and theoretical errors. Additionally, they adopt a binned bias model together with a phenomenological non-perturbative description of the \enquote{Fingers-of-God} effect \cite{Jackson:1971sky}. Their study combines SKA1-MID Band 1 and Band 2, excluding the overlapping redshift range of Band 1 ($0.35 \le z \le 0.49$) as well as low redshifts ($z < 0.1$). While their SKA survey constraints are generally tighter than those presented in Section \ref{sec:results}, our $B_{\ell}$ + Planck analysis yields stronger constraints on the dynamical dark energy parameters in the CPL model. Our results likely benefit from the use of a full-sky approach and a simpler bias model, although they are simultaneously limited by the inclusion of less data. The interplay between these effects will be investigated in future work in order to provide a more direct comparison between the angular and Fourier-space formalisms.

Finally, our forecasts are based on the Fisher matrix formalism and on an idealized description of the experimental performance. The Fisher approach provides the optimal expected uncertainties assuming a Gaussian likelihood and perfect knowledge of the covariance, and we neglected foreground residuals, $1/f$ noise, beam effects, and other instrumental systematics. A fully realistic assessment will require dedicated end-to-end simulations including foregrounds and instrumental effects, combined with a full likelihood analysis.

Despite these limitations, our results show that even within a conservative and theoretically controlled regime, the relativistic angular power spectrum and bispectrum provides a significant gain in cosmological information. Extending the present framework to include redshift bin cross-correlations, improved bias modeling, non-Gaussian covariance, and a consistent treatment of mildly non-linear scales is a natural next step toward fully exploiting the potential of future 21cm intensity-mapping surveys.

\begin{acknowledgments}

R.F.P and A.A.C acknowledge the financial support from CNPq (Brazil) under grants 140696/2022-9 and 102734/2024-0, respectively. This work is supported by National Natural Science Fund of China under Grants Nos. 12175192, and Yangzhou Science and Technology Planning Project in Jiangsu Province of China (YZ2025233). 
\end{acknowledgments}

\bibliography{references}

@article{DESI:2024mwx,
    author = "Adame, A. G. and others",
    collaboration = "DESI",
    title = "{DESI 2024 VI: cosmological constraints from the measurements of baryon acoustic oscillations}",
    eprint = "2404.03002",
    archivePrefix = "arXiv",
    primaryClass = "astro-ph.CO",
    reportNumber = "FERMILAB-PUB-24-0154-PPD",
    doi = "10.1088/1475-7516/2025/02/021",
    journal = "JCAP",
    volume = "02",
    pages = "021",
    year = "2025"
}

@article{Planck:2018vyg,
    author = "Aghanim, N. and others",
    collaboration = "Planck",
    title = "{Planck 2018 results. VI. Cosmological parameters}",
    eprint = "1807.06209",
    archivePrefix = "arXiv",
    primaryClass = "astro-ph.CO",
    doi = "10.1051/0004-6361/201833910",
    journal = "Astron. Astrophys.",
    volume = "641",
    pages = "A6",
    year = "2020",
    note = "[Erratum: Astron.Astrophys. 652, C4 (2021)]"
}

@article{DESI:2016fyo,
    author = "Aghamousa, Amir and others",
    collaboration = "DESI",
    title = "{The DESI Experiment Part I: Science,Targeting, and Survey Design}",
    eprint = "1611.00036",
    archivePrefix = "arXiv",
    primaryClass = "astro-ph.IM",
    reportNumber = "FERMILAB-PUB-16-517-AE",
    month = "10",
    journal = "arXiv preprint",
    year = "2016"
}

@article{DESI:2016igz,
    author = "Aghamousa, Amir and others",
    collaboration = "DESI",
    title = "{The DESI Experiment Part II: Instrument Design}",
    journal = "arXiv preprint",
    eprint = "1611.00037",
    archivePrefix = "arXiv",
    primaryClass = "astro-ph.IM",
    reportNumber = "FERMILAB-PUB-16-518-AE",
    month = "10",
    year = "2016"
}

@article{Amendola:2016saw,
    author = "Amendola, Luca and others",
    title = "{Cosmology and fundamental physics with the Euclid satellite}",
    eprint = "1606.00180",
    archivePrefix = "arXiv",
    primaryClass = "astro-ph.CO",
    doi = "10.1007/s41114-017-0010-3",
    journal = "Living Rev. Rel.",
    volume = "21",
    number = "1",
    pages = "2",
    year = "2018"
}

@article{Euclid:2019clj,
    author = "Blanchard, A. and others",
    collaboration = "Euclid",
    title = "{Euclid preparation. VII. Forecast validation for Euclid cosmological probes}",
    eprint = "1910.09273",
    archivePrefix = "arXiv",
    primaryClass = "astro-ph.CO",
    doi = "10.1051/0004-6361/202038071",
    journal = "Astron. Astrophys.",
    volume = "642",
    pages = "A191",
    year = "2020"
}

@article{LSST:2008ijt,
    author = "Ivezi{\'c}, {\v{Z}}eljko and others",
    collaboration = "LSST",
    title = "{LSST: from Science Drivers to Reference Design and Anticipated Data Products}",
    eprint = "0805.2366",
    archivePrefix = "arXiv",
    primaryClass = "astro-ph",
    reportNumber = "SLAC-PUB-16076",
    doi = "10.3847/1538-4357/ab042c",
    journal = "Astrophys. J.",
    volume = "873",
    number = "2",
    pages = "111",
    year = "2019"
}

@article{LSSTDarkEnergyScience:2018jkl,
    author = "Mandelbaum, Rachel and others",
    collaboration = "LSST Dark Energy Science",
    title = "{The LSST Dark Energy Science Collaboration (DESC) Science Requirements Document}",
    eprint = "1809.01669",
    archivePrefix = "arXiv",
    primaryClass = "astro-ph.CO",
    reportNumber = "FERMILAB-PUB-18-465-A",
    doi = "10.2172/1471560",
    month = "9",
    journal = "arXiv preprint",
    year = "2018"
}

@article{Abdalla:2021nyj,
    author = "Abdalla, Elcio and others",
    title = "{The BINGO project - I. Baryon acoustic oscillations from integrated neutral gas observations}",
    eprint = "2107.01633",
    archivePrefix = "arXiv",
    primaryClass = "astro-ph.CO",
    reportNumber = "TUM-HEP-1323/21",
    doi = "10.1051/0004-6361/202140883",
    journal = "Astron. Astrophys.",
    volume = "664",
    pages = "A14",
    year = "2022"
}

@article{Wuensche:2021dcx,
    author = "Wuensche, Carlos A. and others",
    title = "{The BINGO project - II. Instrument description}",
    eprint = "2107.01634",
    archivePrefix = "arXiv",
    primaryClass = "astro-ph.IM",
    reportNumber = "TUM-HEP-1324/21",
    doi = "10.1051/0004-6361/202039962",
    journal = "Astron. Astrophys.",
    volume = "664",
    pages = "A15",
    year = "2022"
}

@article{Weltman:2018zrl,
    author = "Weltman, A. and others",
    title = "{Fundamental physics with the Square Kilometre Array}",
    eprint = "1810.02680",
    archivePrefix = "arXiv",
    primaryClass = "astro-ph.CO",
    doi = "10.1017/pasa.2019.42",
    journal = "Publ. Astron. Soc. Austral.",
    volume = "37",
    pages = "e002",
    year = "2020"
}

@article{SKA:2018ckk,
    author = "Bacon, David J. and others",
    collaboration = "SKA",
    title = "{Cosmology with Phase 1 of the Square Kilometre Array: Red Book 2018: Technical specifications and performance forecasts}",
    eprint = "1811.02743",
    archivePrefix = "arXiv",
    primaryClass = "astro-ph.CO",
    doi = "10.1017/pasa.2019.51",
    journal = "Publ. Astron. Soc. Austral.",
    volume = "37",
    pages = "e007",
    year = "2020"
}

@article{Costa:2021jsk,
    author = "Costa, Andre A. and others",
    title = "{The BINGO project - VII. Cosmological forecasts from 21 cm intensity mapping}",
    eprint = "2107.01639",
    archivePrefix = "arXiv",
    primaryClass = "astro-ph.CO",
    reportNumber = "TUM-HEP-1329/21",
    doi = "10.1051/0004-6361/202140888",
    journal = "Astron. Astrophys.",
    volume = "664",
    pages = "A20",
    year = "2022"
}

@article{Chen:2019jms,
    author = "Chen, T. and Battye, R. A. and Costa, A. A. and Dickinson, C. and Harper, S. E.",
    title = "{Impact of $1/f$ noise on cosmological parameter constraints for SKA intensity mapping}",
    eprint = "1907.12132",
    archivePrefix = "arXiv",
    primaryClass = "astro-ph.CO",
    doi = "10.1093/mnras/stz3307",
    journal = "Mon. Not. Roy. Astron. Soc.",
    volume = "491",
    number = "3",
    pages = "4254--4266",
    year = "2020"
}

@article{Ivanov:2021kcd,
    author = "Ivanov, Mikhail M. and Philcox, Oliver H. E. and Nishimichi, Takahiro and Simonovi{\'c}, Marko and Takada, Masahiro and Zaldarriaga, Matias",
    title = "{Precision analysis of the redshift-space galaxy bispectrum}",
    eprint = "2110.10161",
    archivePrefix = "arXiv",
    primaryClass = "astro-ph.CO",
    reportNumber = "YITP-21-120, CERN-TH-2021-155",
    doi = "10.1103/PhysRevD.105.063512",
    journal = "Phys. Rev. D",
    volume = "105",
    number = "6",
    pages = "063512",
    year = "2022"
}

@article{Philcox:2021kcw,
    author = "Philcox, Oliver H. E. and Ivanov, Mikhail M.",
    title = "{BOSS DR12 full-shape cosmology: {\ensuremath{\Lambda}}CDM constraints from the large-scale galaxy power spectrum and bispectrum monopole}",
    eprint = "2112.04515",
    archivePrefix = "arXiv",
    primaryClass = "astro-ph.CO",
    doi = "10.1103/PhysRevD.105.043517",
    journal = "Phys. Rev. D",
    volume = "105",
    number = "4",
    pages = "043517",
    year = "2022"
}

@article{Karagiannis:2022ylq,
    author = "Karagiannis, Dionysios and Maartens, Roy and Randrianjanahary, Liantsoa F.",
    title = "{Cosmological constraints from the power spectrum and bispectrum of 21cm intensity maps}",
    eprint = "2206.07747",
    archivePrefix = "arXiv",
    primaryClass = "astro-ph.CO",
    doi = "10.1088/1475-7516/2022/11/003",
    journal = "JCAP",
    volume = "11",
    pages = "003",
    year = "2022"
}

@article{Hall:2012wd,
    author = "Hall, Alex and Bonvin, Camille and Challinor, Anthony",
    title = "{Testing General Relativity with 21-cm intensity mapping}",
    eprint = "1212.0728",
    archivePrefix = "arXiv",
    primaryClass = "astro-ph.CO",
    doi = "10.1103/PhysRevD.87.064026",
    journal = "Phys. Rev. D",
    volume = "87",
    number = "6",
    pages = "064026",
    year = "2013"
}

@article{Umeh:2015gza,
    author = "Umeh, Obinna and Maartens, Roy and Santos, Mario",
    title = "{Nonlinear modulation of the HI power spectrum on ultra-large scales. I}",
    eprint = "1509.03786",
    archivePrefix = "arXiv",
    primaryClass = "astro-ph.CO",
    doi = "10.1088/1475-7516/2016/03/061",
    journal = "JCAP",
    volume = "03",
    pages = "061",
    year = "2016"
}

@article{Maldacena:2002vr,
    author = "Maldacena, Juan Martin",
    title = "{Non-Gaussian features of primordial fluctuations in single field inflationary models}",
    eprint = "astro-ph/0210603",
    archivePrefix = "arXiv",
    doi = "10.1088/1126-6708/2003/05/013",
    journal = "JHEP",
    volume = "05",
    pages = "013",
    year = "2003"
}

@article{Bartolo:2004if,
    author = "Bartolo, N. and Komatsu, E. and Matarrese, Sabino and Riotto, A.",
    title = "{Non-Gaussianity from inflation: Theory and observations}",
    eprint = "astro-ph/0406398",
    archivePrefix = "arXiv",
    reportNumber = "DFPD-04-A-12",
    doi = "10.1016/j.physrep.2004.08.022",
    journal = "Phys. Rept.",
    volume = "402",
    pages = "103--266",
    year = "2004"
}

@article{Kehagias:2015tda,
    author = "Kehagias, A. and Moradinezhad Dizgah, A. and Nore{\~n}a, J. and Perrier, H. and Riotto, A.",
    title = "{A Consistency Relation for the Observed Galaxy Bispectrum and the Local non-Gaussianity from Relativistic Corrections}",
    eprint = "1503.04467",
    archivePrefix = "arXiv",
    primaryClass = "astro-ph.CO",
    doi = "10.1088/1475-7516/2015/08/018",
    journal = "JCAP",
    volume = "08",
    pages = "018",
    year = "2015"
}

@article{DiDio:2016gpd,
    author = "Di Dio, E. and Perrier, H. and Durrer, R. and Marozzi, G. and Moradinezhad Dizgah, A. and Nore{\~n}a, J. and Riotto, A.",
    title = "{Non-Gaussianities due to Relativistic Corrections to the Observed Galaxy Bispectrum}",
    eprint = "1611.03720",
    archivePrefix = "arXiv",
    primaryClass = "astro-ph.CO",
    doi = "10.1088/1475-7516/2017/03/006",
    journal = "JCAP",
    volume = "03",
    pages = "006",
    year = "2017"
}

@article{Jalilvand:2018ikk,
    author = "Jalilvand, Mona and Majerotto, Elisabetta and Durrer, Ruth and Kunz, Martin",
    title = "{Intensity mapping of the 21 cm emission: lensing}",
    eprint = "1807.01351",
    archivePrefix = "arXiv",
    primaryClass = "astro-ph.CO",
    doi = "10.1088/1475-7516/2019/01/020",
    journal = "JCAP",
    volume = "01",
    pages = "020",
    year = "2019"
}

@article{Desjacques:2016bnm,
    author = "Desjacques, Vincent and Jeong, Donghui and Schmidt, Fabian",
    title = "{Large-Scale Galaxy Bias}",
    eprint = "1611.09787",
    archivePrefix = "arXiv",
    primaryClass = "astro-ph.CO",
    doi = "10.1016/j.physrep.2017.12.002",
    journal = "Phys. Rept.",
    volume = "733",
    pages = "1--193",
    year = "2018"
}

@article{Durrer:2020orn,
    author = "Durrer, Ruth and Jalilvand, Mona and Kothari, Rahul and Maartens, Roy and Montanari, Francesco",
    title = "{Full-sky bispectrum in redshift space for 21cm intensity maps}",
    eprint = "2008.02266",
    archivePrefix = "arXiv",
    primaryClass = "astro-ph.CO",
    doi = "10.1088/1475-7516/2020/12/003",
    journal = "JCAP",
    volume = "12",
    pages = "003",
    year = "2020"
}

@article{DiDio:2018unb,
    author = "Di Dio, Enea and Durrer, Ruth and Maartens, Roy and Montanari, Francesco and Umeh, Obinna",
    title = "{The Full-Sky Angular Bispectrum in Redshift Space}",
    eprint = "1812.09297",
    archivePrefix = "arXiv",
    primaryClass = "astro-ph.CO",
    reportNumber = "IFT-UAM/CSIC-18-134, HIP-2018-38/TH",
    doi = "10.1088/1475-7516/2019/04/053",
    journal = "JCAP",
    volume = "04",
    pages = "053",
    year = "2019"
}

@book{Mukhanov:2005sc,
  author    = "Mukhanov, Viatcheslav",
  title     = "{Physical Foundations of Cosmology}",
  publisher = "Cambridge University Press",
  address   = "Cambridge, UK",
  year      = "2005",
  isbn      = "9780521563987"
}

@article{Ma:1995ey,
    author = "Ma, Chung-Pei and Bertschinger, Edmund",
    title = "{Cosmological perturbation theory in the synchronous and conformal Newtonian gauges}",
    eprint = "astro-ph/9506072",
    archivePrefix = "arXiv",
    doi = "10.1086/176550",
    journal = "Astrophys. J.",
    volume = "455",
    pages = "7--25",
    year = "1995"
}

@article{DiDio:2015bua,
    author = "Di Dio, Enea and Durrer, Ruth and Marozzi, Giovanni and Montanari, Francesco",
    title = "{The bispectrum of relativistic galaxy number counts}",
    eprint = "1510.04202",
    archivePrefix = "arXiv",
    primaryClass = "astro-ph.CO",
    reportNumber = "HIP-2015-34-TH",
    doi = "10.1088/1475-7516/2016/01/016",
    journal = "JCAP",
    volume = "01",
    pages = "016",
    year = "2016"
}

@article{Bull:2014rha,
    author = "Bull, Philip and Ferreira, Pedro G. and Patel, Prina and Santos, Mario G.",
    title = "{Late-time cosmology with 21cm intensity mapping experiments}",
    eprint = "1405.1452",
    archivePrefix = "arXiv",
    primaryClass = "astro-ph.CO",
    doi = "10.1088/0004-637X/803/1/21",
    journal = "Astrophys. J.",
    volume = "803",
    number = "1",
    pages = "21",
    year = "2015"
}

@article{Switzer:2013ewa,
    author = "Switzer, E. R. and others",
    title = "{Determination of z{\textasciitilde}0.8 neutral hydrogen fluctuations using the 21 cm intensity mapping auto-correlation}",
    eprint = "1304.3712",
    archivePrefix = "arXiv",
    primaryClass = "astro-ph.CO",
    doi = "10.1093/mnrasl/slt074",
    journal = "Mon. Not. Roy. Astron. Soc.",
    volume = "434",
    pages = "L46",
    year = "2013"
}

@article{Fisher:1935abc,
    author         = "Fisher, R. A.",
    title          = "{The Logic of Inductive Inference}",
    journal        = "J. Roy. Statist. Soc.",
    volume         = "98",
    year           = "1935",
    pages          = "39-82",
    doi            = "10.2307/2342435"
}

@article{Tegmark:1996bz,
    author = "Tegmark, Max and Taylor, Andy and Heavens, Alan",
    title = "{Karhunen-Loeve eigenvalue problems in cosmology: How should we tackle large data sets?}",
    eprint = "astro-ph/9603021",
    archivePrefix = "arXiv",
    doi = "10.1086/303939",
    journal = "Astrophys. J.",
    volume = "480",
    pages = "22",
    year = "1997"
}

@book{Hobson:2009bmc,
    author         = "Hobson, M. P. and Jaffe, A. H. and Liddle, A. R. and Mukherjee, P. and Parkinson, D.",
    title          = "{Bayesian Methods in Cosmology}",
    publisher      = "Cambridge University Press",
    year           = "2009",
    isbn           = "978-0-521-88794-6",
    doi            = "10.1017/CBO9780511596544"
}

@article{Babich:2004yc,
    author = "Babich, Daniel and Zaldarriaga, Matias",
    title = "{Primordial bispectrum information from CMB polarization}",
    eprint = "astro-ph/0408455",
    archivePrefix = "arXiv",
    doi = "10.1103/PhysRevD.70.083005",
    journal = "Phys. Rev. D",
    volume = "70",
    pages = "083005",
    year = "2004"
}

@article{Yadav:2007rk,
    author = "Yadav, Amit P. S. and Komatsu, Eiichiro and Wandelt, Benjamin D.",
    title = "{Fast Estimator of Primordial Non-Gaussianity from Temperature and Polarization Anisotropies in the Cosmic Microwave Background}",
    eprint = "astro-ph/0701921",
    archivePrefix = "arXiv",
    doi = "10.1086/519071",
    journal = "Astrophys. J.",
    volume = "664",
    pages = "680--686",
    year = "2007"
}

@article{Schmit:2018rtf,
    author = "Schmit, Claude J. and Heavens, Alan F. and Pritchard, Jonathan R.",
    title = "{The gravitational and lensing-ISW bispectrum of 21 cm radiation}",
    eprint = "1810.00973",
    archivePrefix = "arXiv",
    primaryClass = "astro-ph.CO",
    doi = "10.1093/mnras/sty3400",
    journal = "Mon. Not. Roy. Astron. Soc.",
    volume = "483",
    number = "3",
    pages = "4259--4275",
    year = "2019"
}

@article{SupernovaSearchTeam:1998fmf,
    author = "Riess, Adam G. and others",
    collaboration = "Supernova Search Team",
    title = "{Observational evidence from supernovae for an accelerating universe and a cosmological constant}",
    eprint = "astro-ph/9805201",
    archivePrefix = "arXiv",
    doi = "10.1086/300499",
    journal = "Astron. J.",
    volume = "116",
    pages = "1009--1038",
    year = "1998"
}

@article{SupernovaCosmologyProject:1998vns,
    author = "Perlmutter, S. and others",
    collaboration = "Supernova Cosmology Project",
    title = "{Measurements of $\Omega$ and $\Lambda$ from 42 High Redshift Supernovae}",
    eprint = "astro-ph/9812133",
    archivePrefix = "arXiv",
    reportNumber = "LBNL-41801, LBL-41801",
    doi = "10.1086/307221",
    journal = "Astrophys. J.",
    volume = "517",
    pages = "565--586",
    year = "1999"
}

@article{Jolicoeur:2020eup,
    author = "Jolicoeur, Sheean and Maartens, Roy and De Weerd, Eline M. and Umeh, Obinna and Clarkson, Chris and Camera, Stefano",
    title = "{Detecting the relativistic bispectrum in 21cm intensity maps}",
    eprint = "2009.06197",
    archivePrefix = "arXiv",
    primaryClass = "astro-ph.CO",
    doi = "10.1088/1475-7516/2021/06/039",
    journal = "JCAP",
    volume = "06",
    pages = "039",
    year = "2021"
}

@article{Weinberg:1988cp,
    author = "Weinberg, Steven",
    editor = "Hsu, Jong-Ping and Fine, D.",
    title = "{The Cosmological Constant Problem}",
    reportNumber = "UTTG-12-88",
    doi = "10.1103/RevModPhys.61.1",
    journal = "Rev. Mod. Phys.",
    volume = "61",
    pages = "1--23",
    year = "1989"
}

@article{Song:2026mqf,
    author = "Song, Yanling and Sang, Yu and Xiao, Linfeng and Zhang, Boyu and Wang, Bin",
    title = "{Forecast on $f(R)$ Gravity with HI 21cm Intensity Mapping Surveys}",
    journal = "arXiv preprint",
    eprint = "2602.05575",
    archivePrefix = "arXiv",
    primaryClass = "astro-ph.CO",
    month = "2",
    year = "2026"
}

@article{Ostergaard:2024brd,
    author = "Ostergaard, Benjamin and da Costa, Andr{\'e} Alencar and Sang, Yu",
    title = "{Cosmological forecasts from the baryon acoustic oscillations in 21 cm intensity mapping}",
    eprint = "2406.04830",
    archivePrefix = "arXiv",
    primaryClass = "astro-ph.CO",
    doi = "10.1142/S021773232550155X",
    journal = "Mod. Phys. Lett. A",
    volume = "40",
    number = "33",
    pages = "2550155",
    year = "2025"
}

@article{Xiao:2021nmk,
    author = "Xiao, Linfeng and Costa, Andre A. and Wang, Bin",
    title = "{Forecasts on interacting dark energy from the 21-cm angular power spectrum with BINGO and SKA observations}",
    eprint = "2103.01796",
    archivePrefix = "arXiv",
    primaryClass = "astro-ph.CO",
    doi = "10.1093/mnras/stab3256",
    journal = "Mon. Not. Roy. Astron. Soc.",
    volume = "510",
    number = "1",
    pages = "1495--1514",
    year = "2021"
}

@article{Baldauf:2016sjb,
    author = "Baldauf, Tobias and Mirbabayi, Mehrdad and Simonovi{\'c}, Marko and Zaldarriaga, Matias",
    title = "{LSS constraints with controlled theoretical uncertainties}",
    journal = "arXiv preprint",
    eprint = "1602.00674",
    archivePrefix = "arXiv",
    primaryClass = "astro-ph.CO",
    month = "2",
    year = "2016"
}

@article{Chan:2016ehg,
    author = "Chan, Kwan Chuen and Blot, Linda",
    title = "{Assessment of the Information Content of the Power Spectrum and Bispectrum}",
    eprint = "1610.06585",
    archivePrefix = "arXiv",
    primaryClass = "astro-ph.CO",
    doi = "10.1103/PhysRevD.96.023528",
    journal = "Phys. Rev. D",
    volume = "96",
    number = "2",
    pages = "023528",
    year = "2017"
}

@article{Biagetti:2021tua,
    author = "Biagetti, Matteo and Castiblanco, Lina and Nore{\~n}a, Jorge and Sefusatti, Emiliano",
    title = "{The covariance of squeezed bispectrum configurations}",
    eprint = "2111.05887",
    archivePrefix = "arXiv",
    primaryClass = "astro-ph.CO",
    doi = "10.1088/1475-7516/2022/09/009",
    journal = "JCAP",
    volume = "09",
    pages = "009",
    year = "2022"
}

@article{Crichton:2021hlc,
    author = "Crichton, Devin and others",
    title = "{Hydrogen Intensity and Real-Time Analysis Experiment: 256-element array status and overview}",
    eprint = "2109.13755",
    archivePrefix = "arXiv",
    primaryClass = "astro-ph.IM",
    doi = "10.1117/1.JATIS.8.1.011019",
    journal = "J. Astron. Telesc. Instrum. Syst.",
    volume = "8",
    pages = "011019",
    year = "2022"
}

@inproceedings{MeerKLASS:2017vgf,
    author = "Santos, Mario G. and others",
    collaboration = "MeerKLASS",
    title = "{MeerKLASS: MeerKAT Large Area Synoptic Survey}",
    booktitle = "{MeerKAT Science}: {On the Pathway to the SKA}",
    eprint = "1709.06099",
    archivePrefix = "arXiv",
    primaryClass = "astro-ph.CO",
    month = "9",
    year = "2017"
}

@article{Jackson:1971sky,
    author = "Jackson, J. C.",
    title = "{Fingers of God: A critique of Rees' theory of primoridal gravitational radiation}",
    eprint = "0810.3908",
    archivePrefix = "arXiv",
    primaryClass = "astro-ph",
    doi = "10.1093/mnras/156.1.1P",
    journal = "Mon. Not. Roy. Astron. Soc.",
    volume = "156",
    pages = "1P--5P",
    year = "1972"
}

@article{Pritchard:2011xb,
  author =	 "Pritchard, Jonathan R. and Loeb, Abraham",
  title =	 "{21-cm cosmology}",
  eprint =	 "1109.6012",
  archivePrefix ="arXiv",
  primaryClass = "astro-ph.CO",
  doi =		 "10.1088/0034-4885/75/8/086901",
  journal =	 "Rept. Prog. Phys.",
  volume =	 75,
  pages =	 086901,
  year =	 2012
}

@article{Battye:2004re,
    author = "Battye, Richard A. and Davies, Rod D. and Weller, Jochen",
    title = "{Neutral hydrogen surveys for high redshift galaxy clusters and proto-clusters}",
    eprint = "astro-ph/0401340",
    archivePrefix = "arXiv",
    doi = "10.1111/j.1365-2966.2004.08416.x",
    journal = "Mon. Not. Roy. Astron. Soc.",
    volume = "355",
    pages = "1339--1347",
    year = "2004"
}

@article{Chang:2007xk,
  author =	 "Chang, Tzu-Ching and Pen, Ue-Li and Peterson,
                  Jeffrey B. and McDonald, Patrick",
  title =	 "{Baryon Acoustic Oscillation Intensity Mapping as a
                  Test of Dark Energy}",
  eprint =	 "0709.3672",
  archivePrefix ="arXiv",
  primaryClass = "astro-ph",
  doi =		 "10.1103/PhysRevLett.100.091303",
  journal =	 "Phys. Rev. Lett.",
  volume =	 100,
  pages =	 091303,
  year =	 2008
}

@article{Loeb:2008hg,
  author =	 "Loeb, Abraham and Wyithe, Stuart",
  title =	 "{Precise Measurement of the Cosmological Power
                  Spectrum With a Dedicated 21cm Survey After
                  Reionization}",
  eprint =	 "0801.1677",
  archivePrefix ="arXiv",
  primaryClass = "astro-ph",
  doi =		 "10.1103/PhysRevLett.100.161301",
  journal =	 "Phys. Rev. Lett.",
  volume =	 100,
  pages =	 161301,
  year =	 2008
}

@article{Sethi:2005gv,
  author =	 "Sethi, Shiv K.",
  title =	 "{HI signal from reionization epoch}",
  eprint =	 "astro-ph/0508172",
  archivePrefix ="arXiv",
  doi =		 "10.1111/j.1365-2966.2005.09485.x",
  journal =	 "Mon. Not. Roy. Astron. Soc.",
  volume =	 363,
  pages =	 "818--830",
  year =	 2005
}

@article{Visbal:2008rg,
  author =	 "Visbal, Eli and Loeb, Abraham and Wyithe, J.Stuart
                  B.",
  title =	 "{Cosmological Constraints from 21cm Surveys After
                  Reionization}",
  eprint =	 "0812.0419",
  archivePrefix ="arXiv",
  primaryClass = "astro-ph",
  doi =		 "10.1088/1475-7516/2009/10/030",
  journal =	 "JCAP",
  volume =	 10,
  pages =	 030,
  year =	 2009
}

\end{document}